\documentclass[preprint,authoryear]{elsarticle}
\usepackage{amssymb,amsmath}
\usepackage{graphicx}
\usepackage{caption}
\usepackage{subcaption}
\usepackage[hyphens,spaces]{url}
\usepackage{color}
\usepackage{xcolor}
\usepackage{listings}
\usepackage{linehighlight}
\usepackage{comment}
\definecolor{codehighlight}{rgb}{0.95,0.8,0.8}
\definecolor{codebackground}{rgb}{0.95,0.95,0.95}
\usepackage{soul}

\newcommand{\domark}{%
  \vbox to 0pt{
    \kern-\dp\strutbox
    \smash{\llap{*\kern1em}}
    \vss
  }%
}

\journal{Engineering Applications of Artificial Intelligence}

\begin{document}
\title{Deep-Learning Driven Noise Reduction for Reduced Flux Computed Tomography}
\author{Khalid L. Alsamadony, Ertugrul U. Yildirim, Guenther Glatz,* Umair bin Waheed, Sherif M. Hanafy \\
* E-mail: guenther@kfupm.edu.sa
}

\begin{abstract}
Deep neural networks have received considerable attention in clinical imaging, particularly with respect to the reduction of radiation risk. Lowering the radiation dose by reducing the photon flux inevitably results in the degradation of the scanned image quality. Thus, researchers have sought to exploit deep convolutional neural networks (DCNNs) to map low-quality, low-dose images to higher-dose, higher-quality images thereby minimizing the associated radiation hazard. Conversely, computed tomography (CT) measurements of geomaterials are not limited by the radiation dose. In contrast to the human body, however, geomaterials may be comprised of high-density constituents causing increased attenuation of the X-Rays. Consequently, higher dosage images are required to obtain an acceptable scan quality. The problem of prolonged acquisition times is particularly severe for micro-CT based scanning technologies. Depending on the sample size and exposure time settings, a single scan may require several hours to complete. This is of particular concern if phenomena with an exponential temperature dependency are to be elucidated. A process may happen too fast to be adequately captured by CT scanning. To address the aforementioned issues, we apply DCNNs to improve the quality of rock CT images and reduce exposure times by more than 60\%, simultaneously. We highlight current results based on micro-CT derived datasets and apply transfer learning to improve DCNN results without increasing training time. The approach is applicable to any computed tomography technology. Furthermore, we contrast the performance of the DCNN trained by minimizing different loss functions such as mean squared error and structural similarity index.
\end{abstract}
\maketitle

\section{Introduction}

Computed tomography has been recognized as an indispensable technology not only in the health care domain but also with respect to industrial applications like reverse engineering \citep{Bartscher2006,Bauer2019}, flaw detection \citep{He2014}, and meteorology to name a few \citep{DeChiffre2014,DuPlessis2016}. The non-destructive nature of CT scanning has also proven to be tremendously valuable in the case of geomaterials, allowing to elucidate transport phenomena in porous media, visualize deformation and strain localization in soils, rocks or sediments, or perform fracture and damage assessment in asphalt, cement and concrete \citep{Alshibli2010}. The three-dimensional data obtained helps to better inform numerical models improving their predictive power and enables delineation of physical properties of the specimen under investigation. Both qualities are particularly valued in the area of digital rock physics \citep{berg2017industrial,alqahtani2020machine}.

The technology has, however, shortcomings, in particular with respect to monitoring dynamic processes. The limitation of prolonged acquisition times is distinctively more severe in case of micro-CT ($\mu$-CT) technologies where it may take several hours for a scan to complete. During acquisition, the object should not physically change - or as little as possible - to allow for meaningful reconstruction of the sinograms. Medical CT scanners, per design, offer significantly shorter acquisition times, mere minutes depending on the sample size. Certain experiments, though, stand to benefit greatly from increased exposure times as noise is decreased. The noise reduction gives rise to better statistics if, for example, porosity is to be estimated in combination with a non-wetting radio contrast agent \citep{Glatz2016}. Similarly, core flood experiments necessitate the presence of a vessel to maintain temperatures and pressures resulting in attenuation of the X-Rays. Again, prolonged acquisition times, in combination with high tube voltages and currents, yield a better image quality. From experience, scanning of a 1-inch long rock specimen using a medical CT scanner operating at maximum tube voltage (from 140 kV to 170 kV), current (200 mA), and exposure time (four seconds per slice), requires up to 30 minutes between scans to allow for the X-Ray tube to cool down. Conversely, certain reactive processes happen rather rapidly, mandating low exposure times if the dynamics are to be captured. This is particularly true for high-temperature experiments given the exponential dependency of the reaction rate on heat \citep{Glatz2018, Boigne2020}.

The medical domain of low dose computed tomography (LDCT) seeks to reduce the exposure time in an effort to minimize the radiation risk \citep{mccollough2009strategies}. Lowering the flux by reducing the exposure time, tube peak voltages, and currents will inevitably decrease the image quality and, thereby, the diagnostic value \citep{goldman2007principles}. Photon emission from the X-Ray source is modeled as a Poisson process \citep{macovski_medical_1983} and photon starvation at the detector gives rise to Poisson noise \citep{barrett2004artifacts,Gravel2004}. Additional noise is introduced during the quantization of the signal and in the form of electronic noise \citep{Diwakar2018}. Naturally, researchers sought to reduce artifacts employing improved algorithms during the reconstruction of the 3D data from the projections/sinograms \citep{willemink2019evolution} or post-reconstruction. Generally, the latter approach is more common given that the raw CT data is often not accessible, especially in the case of a medical CT system \citep{nishio2017convolutional,chen2017low}.

Conventional signal processing techniques require a good understanding of the underlying nature of noise to optimize the filter design. Noise statistics, to guide the filter model, may be collected experimentally but, from experience, this constitutes a rather laborious process.

Recently, deep convolutional neural networks (DCNNs) have been successfully applied to map low-quality, low-dose images to higher-dose, higher-quality images \citep{Chen2017,Kang2017}. In this paper, we seek to build on this general approach and apply it to computed tomography scanning of geomaterials for the following reasons. First, the reduction of acquisition time allows for an increase in the temporal resolution. Consequently, experiments previously deemed out of reach due to the associated dynamics can now be entertained. Second, with respect to digital rock physics, the accuracy of the estimated rock properties strongly depend on the image quality \citep{bazaikin2017effect,liu2018critical,guan2019effects}. Third, high quality images are a prerequisite for resolution enhancement techniques \citep{papari2016fast,wang2019ct,da2019enhancing}. In addition, a reduction in scan time will contribute to an increase in the lifetime of the X-Ray tube (medical CT) and the filament ($\mu$-CT), respectively.

Using artificial rock CT images obtained from simulated parallel-beam projections, \cite{pelt2018improving} obtained promising results with respect to improved image quality by means of DCNNs. The work presented in this paper seeks to extend the DCNNs filtering concept including results not only for synthetic data created using the ASTRA toolbox \citep{VanAarle2015} but also for actual $\mu$-CT data generated by a FEI Heliscan microCT operating with a cone-beam. Importantly, we do not only aim to reduce scanning time but seek to improve the quality of the reconstructed images compared to the high dose training images, simultaneously.

In short, in this paper, we study two deep learning architectures for improving degraded rock images resulting from reduced exposure time $
\mu$-CT scans. Furthermore, we investigate the applicability of transfer learning to minimize the number of training images needed. In addition, we explore the impact of mean-squared error (MSE) and structural similarity index measure (SSIM) loss functions on the reconstructed image quality. While both loss functions are capable of considerably improving the respective quality metrics, PSNR and SSIM, they tend to emphasize different structural features. These findings are crucial for improving digital rock physics applications where rock properties, like porosity and permeability, are to be estimated from computed tomography data only.

\section{Methodolgy}
Convolutional neural networks (CNNs) constitute a subset of artificial neural networks (ANNs), heavily relying on digital filter operations (kernel/convolution matrix), where the weights of the filters are informed during the training process to minimize a particular loss metric between the predictions and the training (true) samples. CNNs are especially suitable for computer vision applications given that the filters can capture the spatial relation between individual pixels or image elements. 

The main components of CNNs are convolution layers and activation functions. Convolution layers consists of filters that slide across the input feature map (e.g., image). Each element of a filter is multiplied by the overlapping element of the input feature map and subsequently summed to yield one element of the output feature map. This operation is followed by activation functions to add non-linearity empowering a CNN to learn the complex relationship between inputs and their corresponding labels. A common activation function is the rectified linear unit (ReLU), designed to set negative values to zero, and linearly map inputs to outputs in case of positive values. 

\subsection{Details of Network Architectures Investigated}
\begin{figure}
    \includegraphics[width=\linewidth]{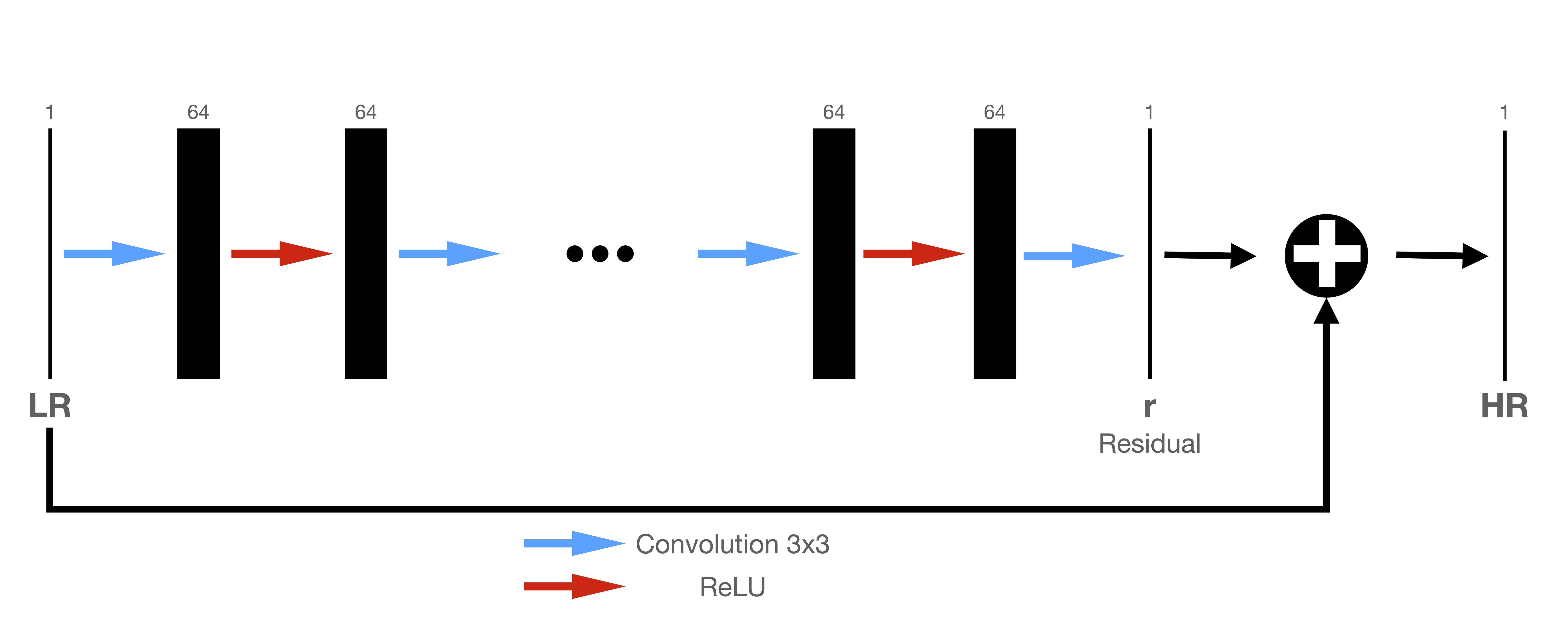}
  \caption{VDSR architecture showing cascaded pair of layers. The input is a low-resolution image, or a noisy image in our case, which goes through layers and gets transformed to a high-resolution or denoised image. The convolutional layers use 64 filters each.}
  \label{fig:vdsr}
\end{figure}

For the transfer learning aspect of this work, we take advantage of the \emph{very deep super resolution} (VDSR) architecture and the associated pretrained VDSR by \cite{kim2016accurate}. Both were obtained from the MathWorks website \citep{mathworks}. The VDSR architecture consists of 20 weighted convolution layers followed by a ReLU. Each convolution layer, except the final and the input layer, accommodate 64 filters of size 3$\times$3. Figure~\ref{fig:vdsr} shows the network architecture.

\begin{figure}
    \includegraphics[width=\linewidth]{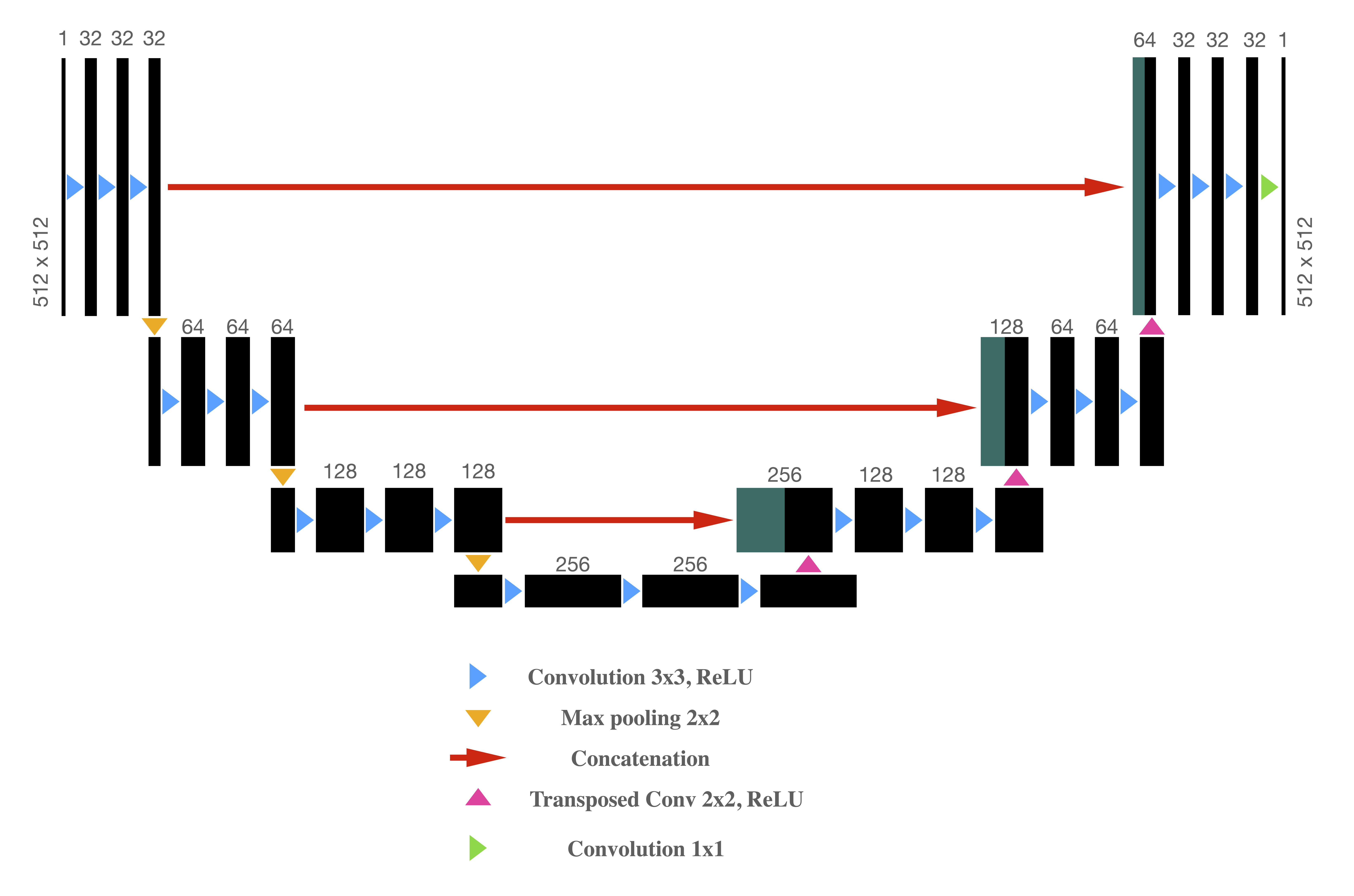}
  \caption{Architecture of the proposed DCNN (U-Net) which is based on a residual encoder/decoder structure. Each black box represents a feature map. The number of channels is denoted at the top of each box. The input and output images have the same size (height and width) which is indicated at the sides of the first and last box. Dark green boxes represent copied feature maps from the encoder block. The arrows state different operations.}
  \label{fig:dcnn}
\end{figure}

 The second architecture investigated constitutes a deep convolutional neural network (DCNN) based on a residual encoder/decoder architecture, which is known as U-Net network. The encoder or feature extractor component consists of three blocks with each block comprising three consecutive convolution layers where all layers are followed by a ReLU activation function. At the end of each block, a sample-based discretization process in form of max pooling operation is executed to reduce the size of the feature map. Similarly, the decoder incorporates a transposed convolution layer followed by three consecutive convolution layers to be terminated by ReLU activation functions. In addition, skip connections between each encoder block and its corresponding decoder block are included to concatenate the output of the transposed convolution layers with the feature map from each encoder block. All convolution layers, except the last layer, are of size 3$\times$3. In the encoder part, the number of filters increases for each block (32, 64, and 128), sequentially. Equivalently, in the decoder part, the number of filters decreases for each block (128, 64, and 32), sequentially. The network configuration is outlined in Figure~\ref{fig:dcnn}.

\subsection{Loss Functions}
The mean squared error (MSE) is a commonly used loss function for image restoration tasks and is defined as follows:

\begin{equation}
\mathcal{L}^{\text{MSE}} \left(P\right)=\frac{1}{N}{\sum_{p \in P}\left[ t\left(p\right)-r\left(p\right)\right]}^2,
\label{eqn:mse}
\end{equation}
where $p$ constitutes the index of the pixel in patch $P$, $t(p)$ is the pixel value in the trained patch, and $r(p)$ corresponds to the pixel value of the reference image, and $N$ is the number of pixels in a given patch.  \citep{zhao2016loss}.

The structural similarity index measure (SSIM), however, is often regarded as a more pragmatic metric for evaluating image quality, particularly with respect to human visual perception \citep{wang2004image}. For a pixel $p$, the SSIM is given as follows:

\begin{equation}
\text{SSIM}(p)=\frac{(2\mu_r\mu_t+c_1)(2\sigma_{tr}+c_2)}{(\mu_r^2+\mu_t^2+c_1)(\sigma_r^2+\sigma_t^2+c_2)},
\label{eqn:ssim}
\end{equation}
where $\mu$ and $\sigma^2$ reflect the average and variance of the training and reference patch, respectively, and $\sigma_{tr}$ is the associated covariance. $c_1$ and $c_2$ are constants required to avoid division with a weak denominator and partially depend on the dynamic range of the pixels. Given that the SSIM ranges from -1 to 1, with 1 being indicative that the training image is identical to the reference image, the loss function needs to be written as follows: 
\begin{equation}
\mathcal{L}^{\text{SSIM}}(P) =\frac{1}{N} \sum_{p \in P}1 - \text{SSIM}(p).
\label{eqn:eq3}
\end{equation}  

For training of the VDSR, and the pretrained VDSR, an image patch size of 41$\times$41 with 128 patches per image was utilized. The Adam optimizer was configured for a learning rate of 0.0001, 5 epochs, and a mini-batch size of 32. For this case we only applied the MSE loss function as defined in Eq.\ref{eqn:mse}.

Both loss functions were exploited to train the DCNN (U-Net) employing an image patch size of 512$\times$512, 1600 training images, 400 testing images, the Adam optimizer configured for a learning rate of 0.0001, 50 epochs, and a mini-batch size of 8.

Naturally, for both architectures, the low quality/low exposure images served as input to be trained on the corresponding high quality/high exposure scans.

\subsection{Data Acquisition Details}
Using a FEI Heliscan microCT, configured to perform 1800 projections per revolution at a tube voltage of 85 kV and a current of 72 mA, two datasets were acquired at an exposure time of 0.5 seconds and 1.4 seconds, respectively. As mentioned above, an increased exposure time translates to a greater image quality given that more photons are collected at the detector thereby decreasing the noise. Henceforth, we refer to the scans collected at 1.4 seconds as high quality images and 0.5 seconds data as low quality images. The scanned specimen was of carbonate origin measuring 1.5 inches in diameter and about 2 inches in length. The sample geometry dictated a minimum voxel size of about 14 $\mu m$.

During all scans, an approximately 100 $\mu m$ thick aluminum sheet was mounted at the tungsten target window to soften the X-Rays in an effort to minimize beam hardening artefacts. The amorphous-silicon, large-area, digital flat-panel detector with 3072$\times$3072 pixels, is capable of supporting a pixel array of 9 mega-pixels with a dynamic range of 16 bits. The effective scan resolution was 2884$\times$2884 pixels. To accommodate the network architecture, the individual slices needed to be split into tiles of size 512$\times$512 pixels and the gray-scale values were normalized to a range between zero and one. In a first order approximation, the gray values may be interpreted as density values where brighter areas are indicative of greater density and darker areas of lower density. Hence, pore space is represented by shades of black (see e.g., Figure \ref{fig:high_and_low_exposure}).

\subsection{ASTRA Toolbox}
As will be shown later, the images predicted by the network are of significantly greater quality than the training images collected at an exposure time of 1.4 seconds. Consequently, it became necessary to create an artificial case based on the images predicted by the network to verify that the architecture is indeed predicting the ground truth.

The ASTRA toolbox is an open-source software for tomographic projections and reconstruction, available for MATLAB$^{\copyright}$ and Python \citep{VanAarle2015,van2016fast}. Throughout this work, MATLAB$^{\copyright}$ 2020a in combination with the ASTRA toolbox V1.9 was utilized. The cone beam projection module in the ASTRA toolbox offers the following three reconstruction algorithms: FDK by \citeauthor{Feldkamp1984}, simultaneous iterative reconstruction technique (SIRT) by \citeauthor{Gilbert1972} and conjugate gradient least squares (CGLS) by \citeauthor{Frommer1999}. 

The toolbox allowed us to simulate artificial low and high exposure time images based on images predicted by the network. Given that the toolbox does not model noise sources and, effectively, assumes a perfect detector, varying degrees of Poisson noise were added to the projections to simulate the physical process at the detector. Subsequently, the sinograms were reconstructed by means of the FDK algorithm to yield artificial low and high exposure time scans. To summarize, at this point the following image series are available:

\begin{enumerate}
    \item A 0.5 seconds exposure time series obtained from the $\mu$-CT, constituting the low quality data (see e.g. right-hand-side in Fig. \ref{fig:high_and_low_exposure}).
    \item A 1.4 seconds exposure time series obtained from the $\mu$-CT, representing the high quality data (see e.g. left-hand-side in Fig. \ref{fig:high_and_low_exposure}).
    \item The images predicted by the network which are of greater quality compared to the training images (labels). These images serve as the ground truth.
    \item An artificial low quality image series, derived from predicted images using the ASTRA toolbox mimicking the results obtained from the $\mu$-CT at an exposure time of 0.5 seconds.
    \item An artificial higher quality image series, delineated from predicted images using the ASTRA toolbox resembling the results obtained from the $\mu$-CT at an exposure time of 1.4 seconds.
\end{enumerate}

The artificially created series was used to validate the predictive power of the trained network as detailed in Section~\ref{astra_section}.

\section{Results}

In this section, we benchmark the proposed DCNNs to restore low quality $\mu$-CT images as a result of reduced exposure times. We begin by highlighting the problem and its adverse consequences on the scanned image quality. Next, we explore the applicability of transfer learning to help expedite the training of the PVDSR network.

Exploiting a pre-trained VDSR network we substantiate that optimal performance can be obtained faster than relying on a randomly initialized VDSR network. In addition, we also compare the reconstruction performance of different loss functions including MSE and SSIM. Finally, we prove the efficacy of the DCNNs by testing it against simulated low and high exposure images from the ASTRA toolbox~\citep{VanAarle2015}. 

\subsection{Reduced Exposure}
In the context of rock imaging, or imaging of materials in general, the reduction of exposure time offers three main advantages.

Firstly, $\mu$-CT scanning, if offered as a commercial service is, from experience, charged on an hourly basis ranging from hundreds to thousands of dollars per hour. Evidently, high quality scans necessitate a longer exposure time consequently being more costly. Thus, a decrease in scan time while maintaining image quality is beneficial to both parties: it allows the provider to offer the service to the client at a reduced cost, and, at the same time, increase the throughput.

Secondly, any reduction in exposure time will results in a more economic use of the filament life time. Generally, a single filament costs about $700$ to $1,000$ dollars and is rated for about 300 working hours. Assuming the particular scan time reduction achieved in this work, roughly $60\%$, the filament life time may be increased up to 480 working hours. In addition, as will be shown later, the network also performs exceedingly well at denoising the image without the need for user intervention.

Thirdly, and most importantly, a reduction in exposure time renders the technology available to elucidate processes previously out of reach due to the associated dynamics.

\begin{figure}[htb]
  \includegraphics[width=\linewidth]{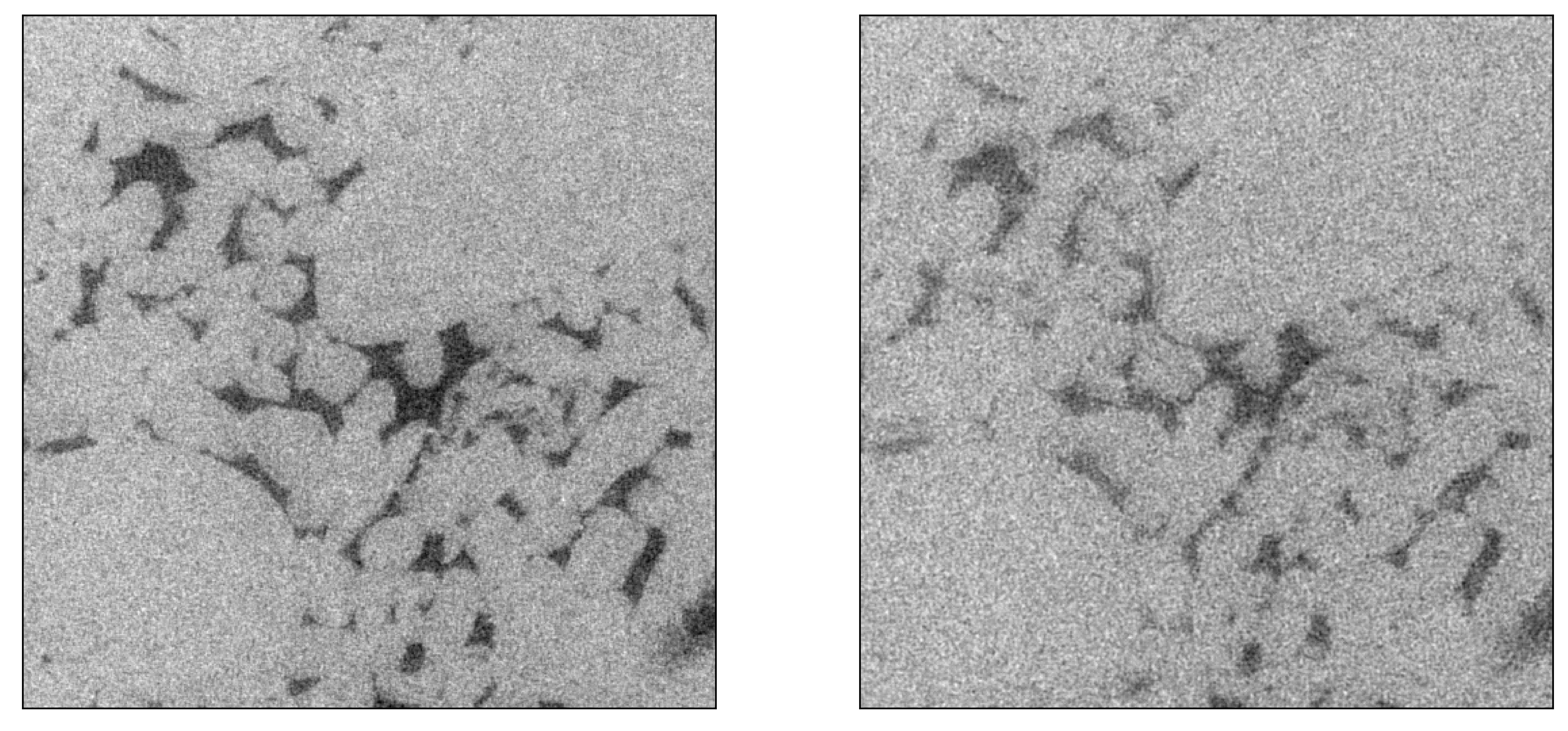}
  \caption{High exposure time, 1.4 seconds, CT image (left) and low exposure time, 0.5 seconds, CT image (right) of a carbonate rock sample where dark colors are indicative of pore space. Evidently, a reduced exposure time results in an increased noise level owing to the photon starvation at the detector.}
  \label{fig:high_and_low_exposure}
\end{figure}

As mentioned before, a reduction in exposure time increases the noise level due to photon starvation at the detector. For example, Figure \ref{fig:high_and_low_exposure} illustrates how lower exposure time CT data (0.5 seconds) yields a considerably noisier image compared to a 1.4 second exposure time scan. 

The noise present in the low exposure, and, noticeably, also in the high exposure image, is problematic if rock properties like porosity and permeability are to be estimated. Accurate porosity values strongly rely on the ability to precisely distinguish between the solid phase and the pore space. With respect to Figure \ref{fig:high_and_low_exposure} this can constitute a daunting task, especially for the low exposure time case. Commonly, a median or smoothing filter followed by, for example, a histogram or watershed based segmentation is applied \citep{Avsar2017}.

Estimation of permeability is significantly more involved as it necessitates a computational fluid dynamics study on the segmented data \citep{Mostaghimi2013}. In addition, permeability is particularly dependent on fine scale features and mineralogy in case of wetting fluids. 


To address the aforementioned challenges, we sought to train DCNNs to denoise low exposure $\mu$-CT images without the need for expert knowledge with respect to filter design.


\subsection{Transfer Learning}
While DCNNs have shown remarkable performance for a myriad of scientific problems, they are well-known for being data- and resource-intensive due to the large number of trainable parameters. Lack of training samples or computational resources may hurt the performance of these networks in either of these situations. A pragmatic approach to address this issue is to take advantage of transfer learning, a machine learning technique seeking to apply previously gained knowledge to speed up finding the solution to a different yet related problem.

In this particular study, we explore the applicability of transfer learning using the VDSR network, as illustrated in Figure~\ref{fig:vdsr}, and compare the reconstruction performance of the VDSR network initialized as per \cite{DBLP:journals/corr/HeZR015} with a pre-trained VDSR network by minimizing the MSE as defined in Equation \ref{eqn:mse}. We train both the pre-trained VDSR network and the VDSR network for a range of number of training images, starting with 50 training images up to a maximum of 300 training images. For each particular number of training images, we measure the reconstruction performance of the two networks by comparing the average SSIM and peak signal-to-noise ratio (PSNR) values derived from 400 test images. 

\begin{figure}[htb]
  \includegraphics[width=\linewidth]{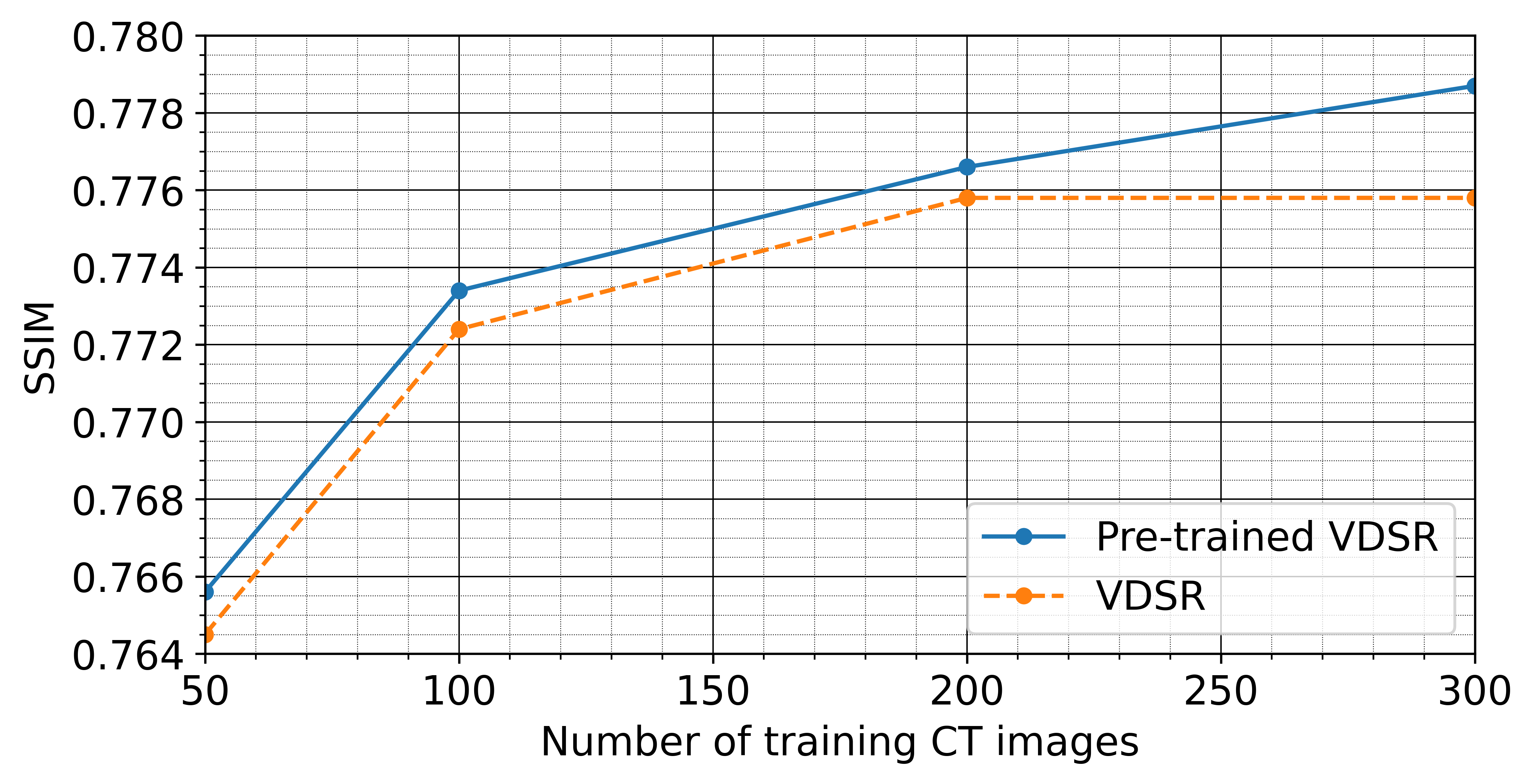}
  \caption{Summary plot of the average SSIM values employing 400 test images as predicted by the pre-trained VDSR network and a VDSR initialized following the approach of \cite{DBLP:journals/corr/HeZR015}. From 50 to 200 training images, both networks show similar performance gains with respect to the corresponding SSIM values. After 200 training images, however, the VDSR to further increase the SSIM value compared to the pre-trained network. In general, however, the pre-trained VDSR yields greater SSIM values for all cases.}
  \label{fig:ssim_comparison_plot}
\end{figure}

\begin{figure}[htb]
    \includegraphics[width=\linewidth]{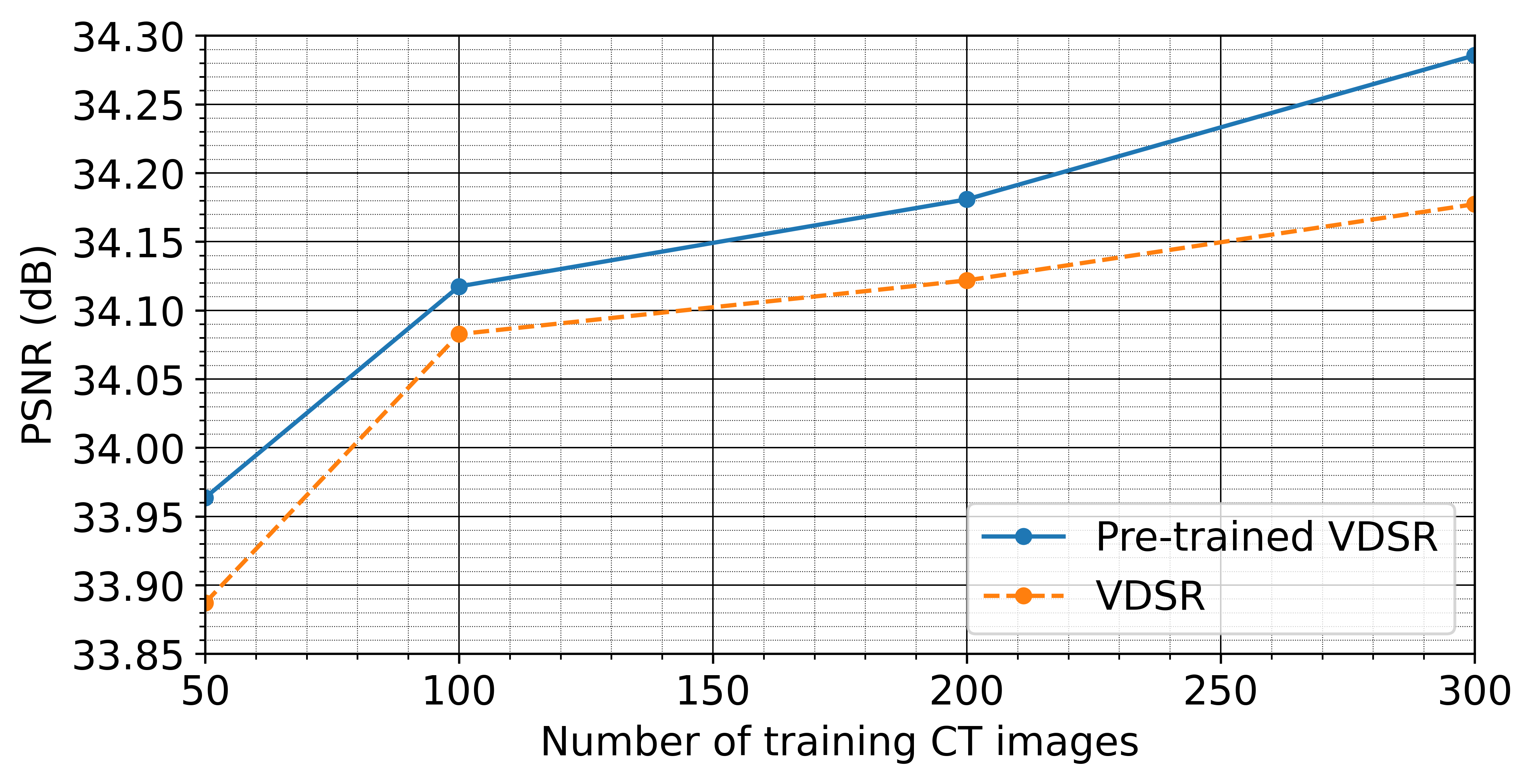}
  \caption{Summary plot of the average PSNR values employing 400 test images as predicted by the pre-trained VDSR network and the VDSR. From 50 to 100 training images the VDSR shows a slightly better improvement compared to the pre-trained VDSR network. This advantage diminishes as the number of training images increases. Similar to the SSIM plot shown in Figure~\ref{fig:ssim_comparison_plot}, the pre-trained VDSR yields greater SSIM values for all cases.}
  \label{fig:psnr_comparison_plot}
\end{figure}

From Figures~\ref{fig:ssim_comparison_plot} and~\ref{fig:psnr_comparison_plot}, we observe that the pre-trained VDSR network always yields a better overall reconstruction performance for a given number of training images. This holds true for both considered metrics to quantify reconstruction quality (SSIM and PSNR), demonstrating the inherent advantage of transfer learning.

\begin{figure}[htb]
    \includegraphics[width=\linewidth]{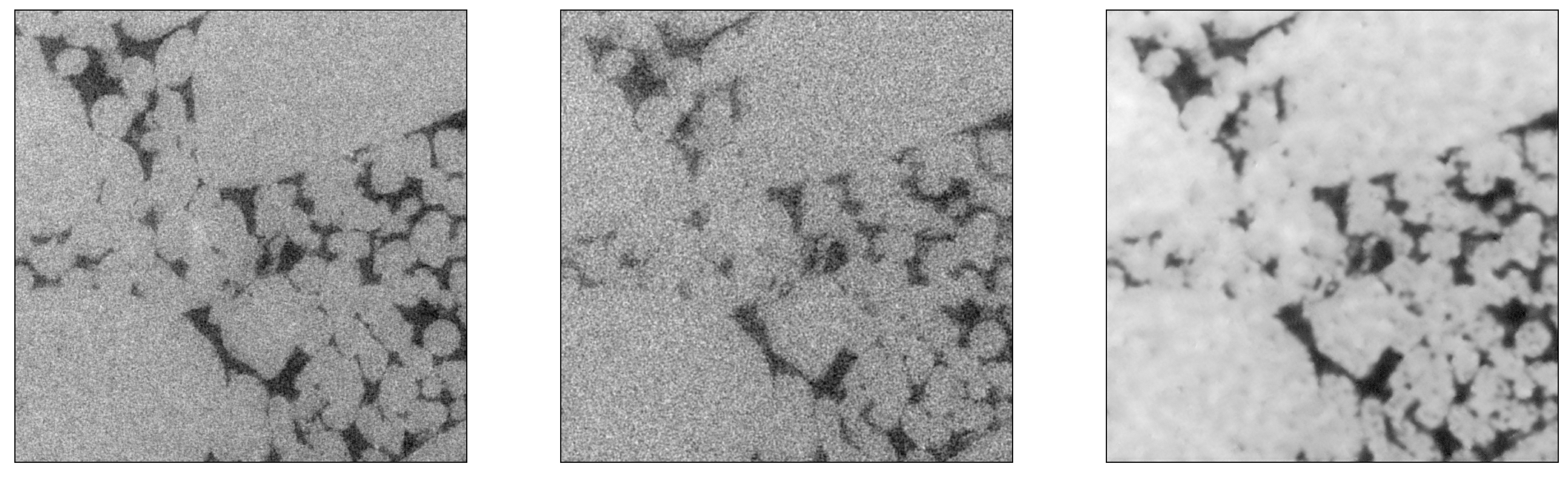}
  \caption{From left to right: High exposure (reference) image, low exposure image (SSIM=0.54, PSNR=23 dB), and denoised image (SSIM=0.78, PSNR=34 dB) using the pre-trained VDSR network.}
  \label{fig:pre_trained_VDSR_example}
\end{figure}

Figure~\ref{fig:pre_trained_VDSR_example} exemplifies the reconstruction quality achieved by the pre-trained VDSR network, based on 300 training images, improving the SSIM and PSNR values of the low exposure image from 0.54 and 23 dB to 0.78 and 34 dB, respectively.

Examining the predicted image in Figure~\ref{fig:pre_trained_VDSR_example}, it should be noted that it is indeed of greater quality (less noisy) than the high quality reference image or training label. The noise is greatly reduced and the grain boundaries show a sharper delineation. This is somewhat surprising given that from a conventional signal processing point of view both edges and noise constitute high frequency content. Often, filters designed to remove high frequency content are often found to smear out edges and subtle details \citep{Lee1981}. Granted, median filters, or filters utilizing local statistics, in general, perform well in preserving them yet it is remarkable that the network learned to differentiate between discontinuities in form of edges and noise. This particular aspect is addressed in more detail in Section \ref{astra_section}.

\subsection{Loss Functions}

Proper selection of the loss, objective, or fitness function is crucial in guiding the learning process of the network. The MSE loss function, as defined in Equation \ref{eqn:mse}, is a preferred metric to optimize the weights owing to its simplicity and well-behavedness with respect to gradient calculations. Notably, minimization of the MSE indirectly maximizes the PSNR.

\begin{figure}
\centering
    \includegraphics[width=\linewidth]{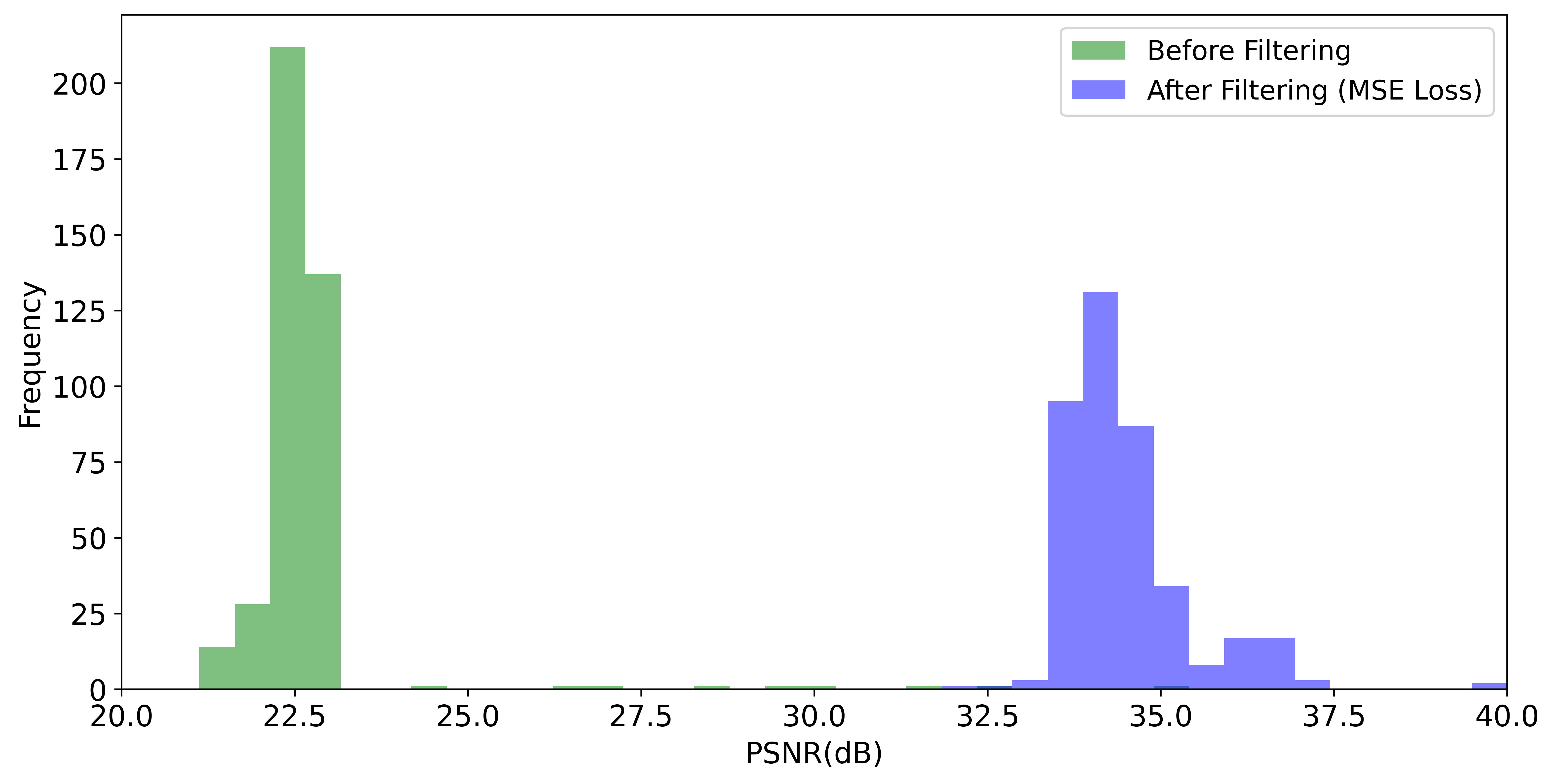}
  \caption{Histogram of the PSNR values obtained for the 400 test images. “Before filtering” refers to the low exposure scans. “After filtering” refers to the DCNN (U-Net) denoised scans where the network as optimized with respect to the MSE loss function.}
  \label{fig:psnr_histogram_mse_loss}
  
      \includegraphics[width=\linewidth]{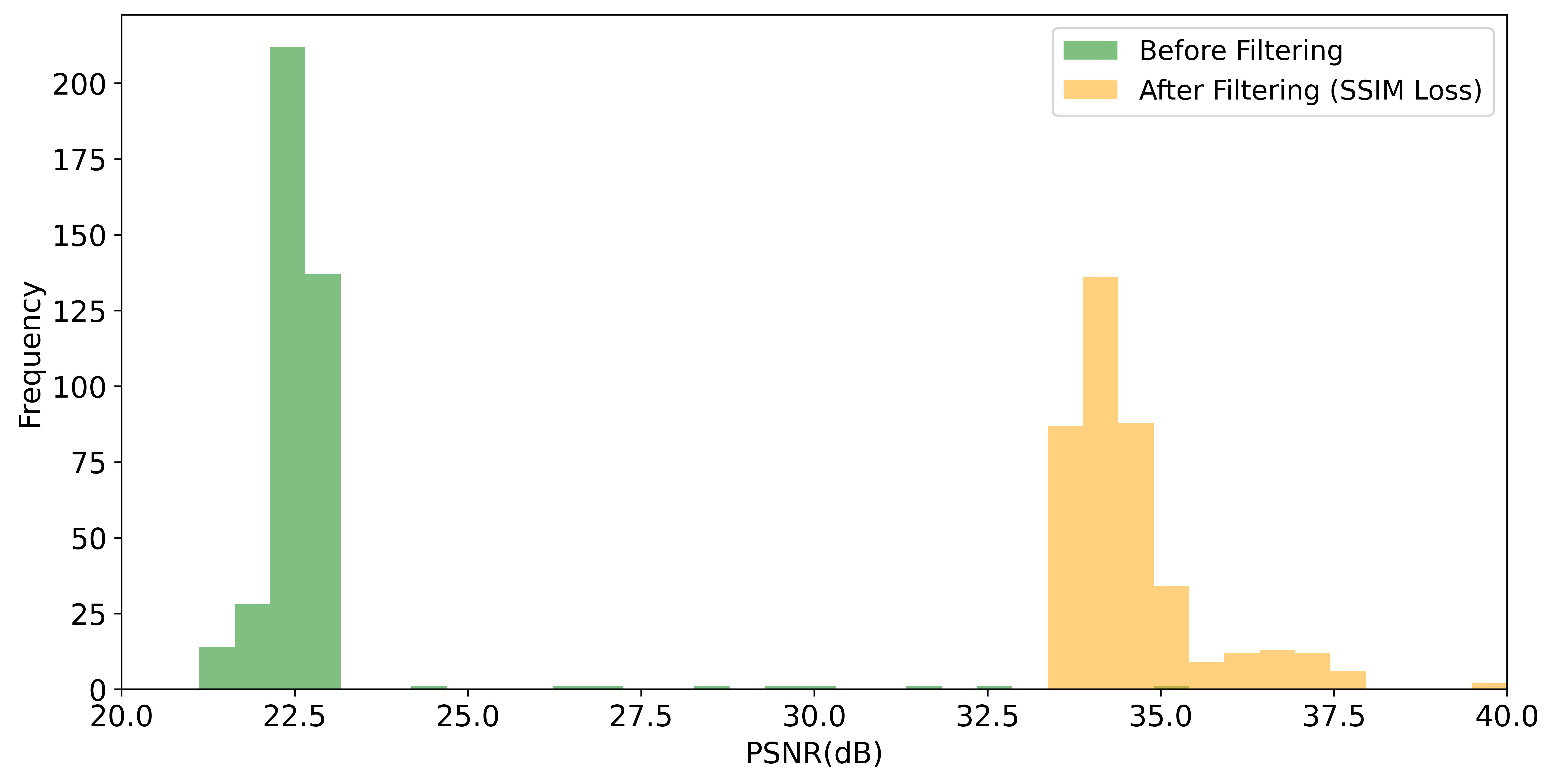}
  \caption{Histogram of the PSNR values obtained for the 400 test images. “Before filtering” refers to the low exposure scans. “After filtering” refers to the DCNN (U-Net) denoised scans where the network was optimized with respect to the SSIM loss function.}
  \label{fig:psnr_histogram_ssim_loss}
\end{figure}

\begin{figure}
      \includegraphics[width=\linewidth]{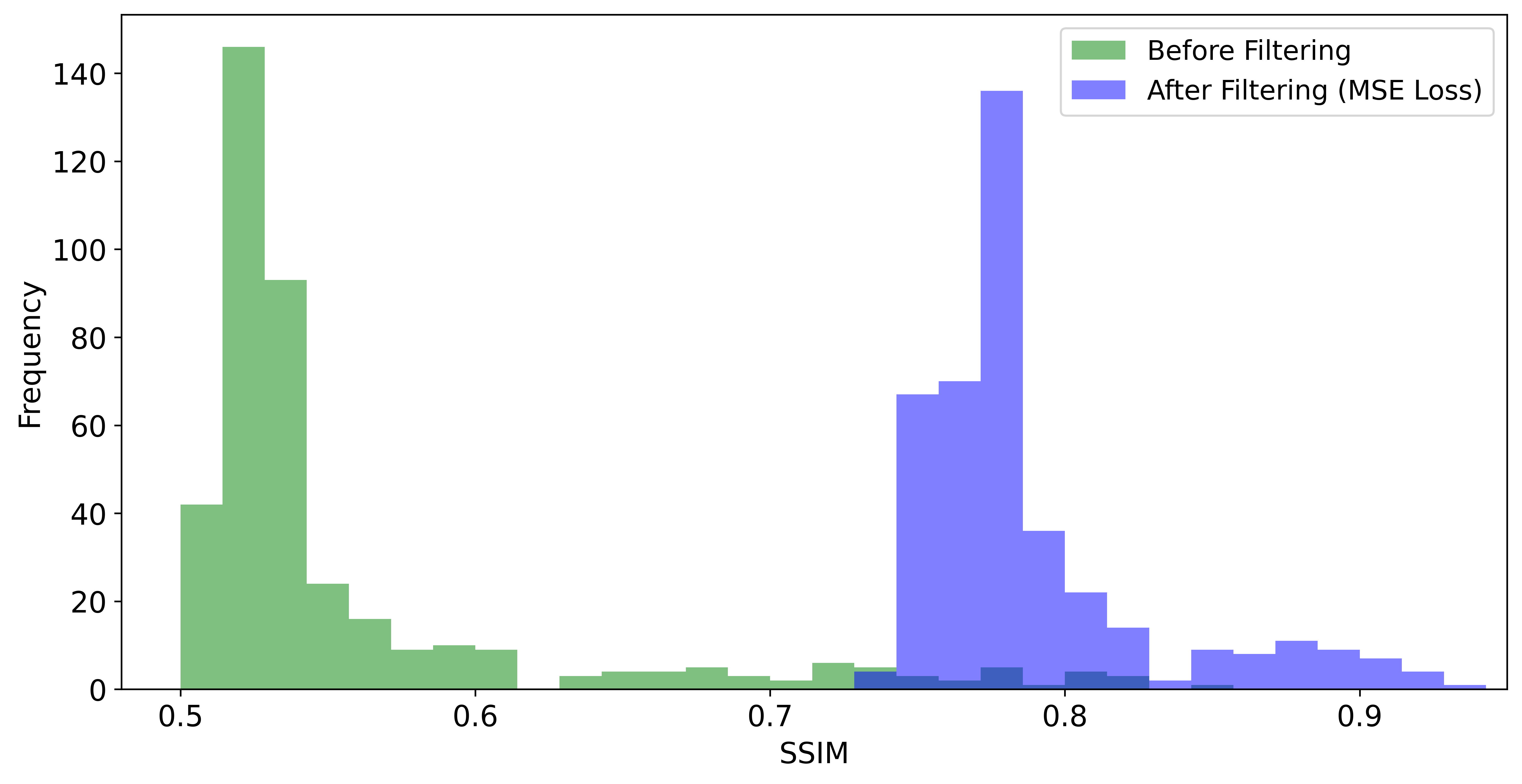}
  \caption{Histogram of the SSIM values obtained for the the 400 test scans. “Before filtering” refers to the low exposure images. “After filtering” refers to the DCNN (U-Net) denoised scans where the network was optimized with respect to the MSE loss function.}
  \label{fig:ssim_histogram_mse_loss}
  
      \includegraphics[width=\linewidth]{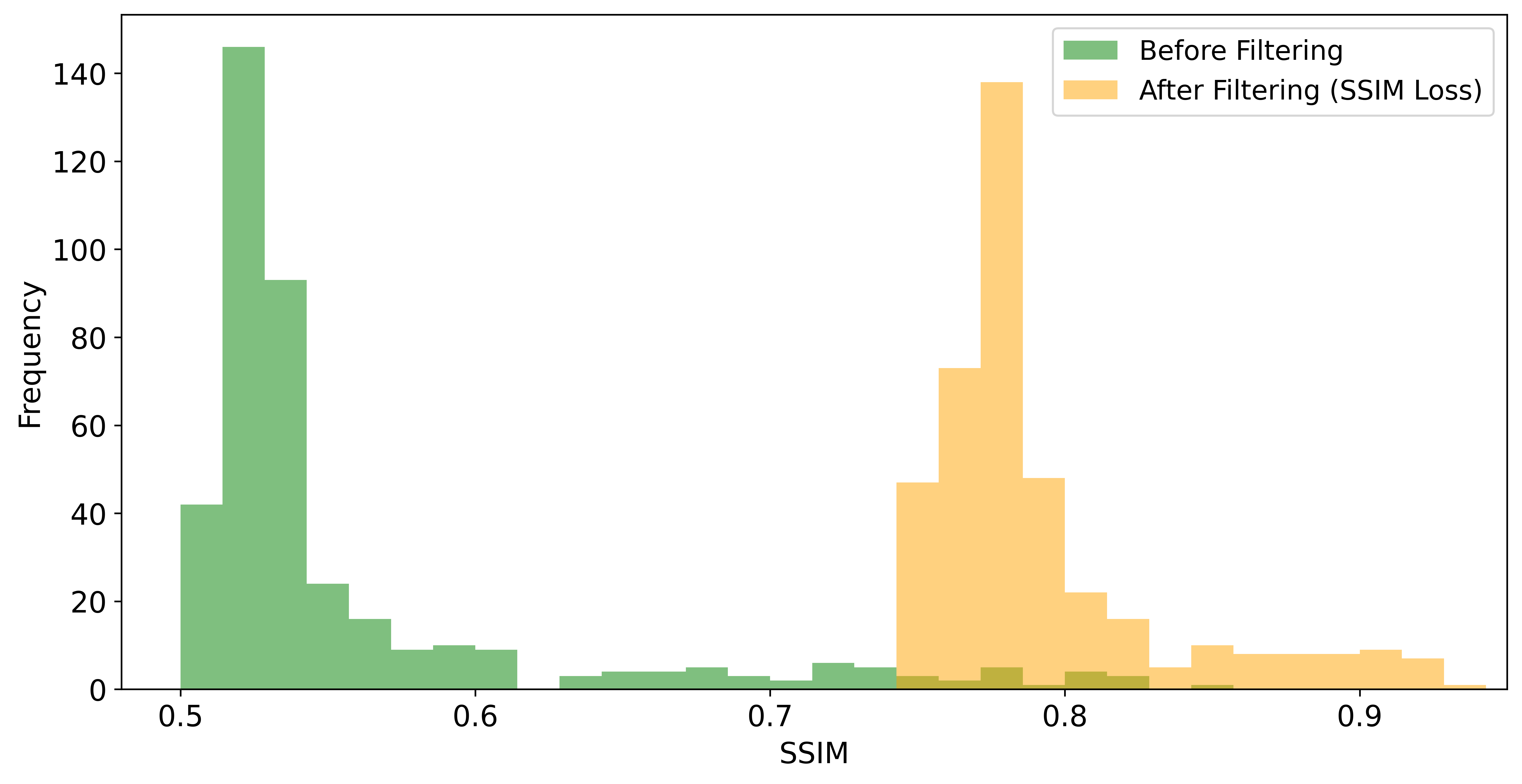}
  \caption{Histogram of the SSIM values obtained for the the 400 test scans. “Before filtering” refers to the low exposure images. “After filtering” refers to the DCNN (U-Net) denoised scans where the network was optimized with respect to the SSIM loss function.}
  \label{fig:ssim_histogram_ssim_loss}  
\end{figure}

Given the particular problem of image prediction, we seek to compare the impact of the MSE loss function against the SSIM loss function, as defined in Equation \ref{eqn:ssim}, on the image quality metrics PSNR and SSIM, respectively. For this purpose, we train the U-net-derived DCNN on 1600 training images for each metric. The trained networks were benchmarked using 400 test images. In general, we obtain considerable improvements for both the SSIM and PSNR values of the reconstructed images as shown in Figures~\ref{fig:psnr_histogram_mse_loss}-\ref{fig:ssim_histogram_ssim_loss}. 

\begin{figure}[htb]
    \includegraphics[width=\linewidth]{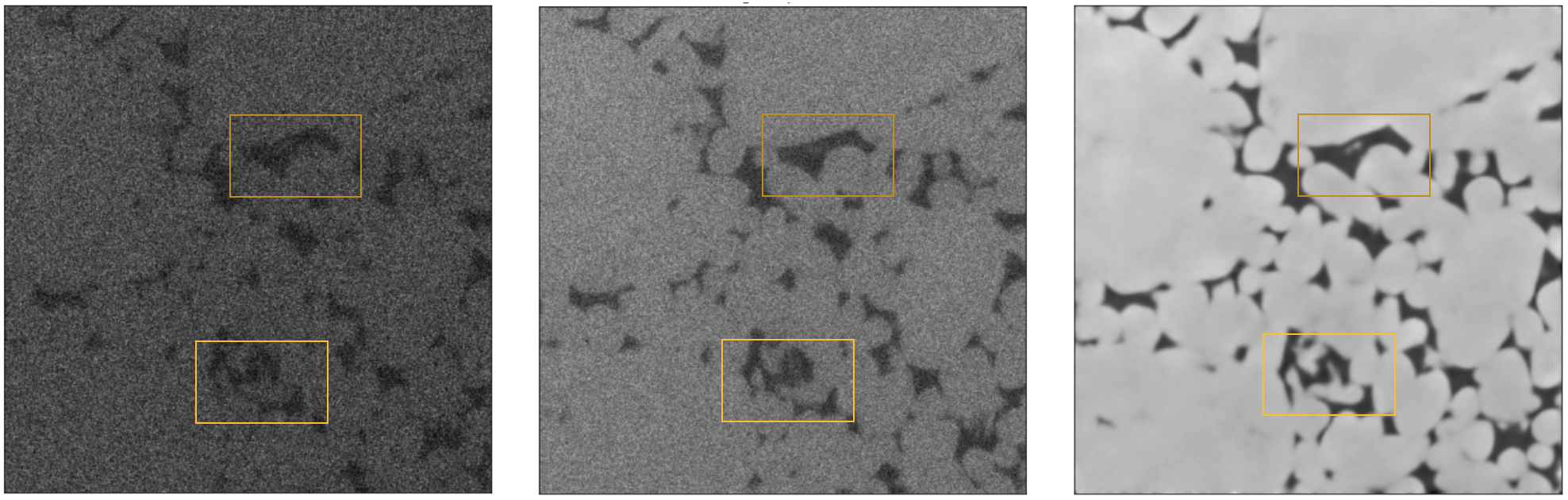}
  \caption{The image on the left is an example of a low exposure time slice with the image in the center being the high exposure time equivalent. The image on the far right is the reconstruction based on the SSIM optimized DCNN (U-Net). The DCNN performs remarkably well in reconstructing fine scale features, barely visible even in case of a longer exposure time.}
  \label{fig:details_example}
\end{figure}

For the MSE optimized network, the PSNR increased, on average, from about 22.6 dB to 34.5 dB (see Figure \ref{fig:psnr_histogram_mse_loss}), and the SSIM from 0.56 to 0.79 (see Figure \ref{fig:ssim_histogram_mse_loss}). In case of the SSIM optimized network the PSNR increased, on average, to 34.6 dB (see Figure \ref{fig:psnr_histogram_ssim_loss}), and the SSIM to 0.79 (see Figure \ref{fig:ssim_histogram_ssim_loss}).

\begin{figure}
     \centering
     \begin{subfigure}[b]{\textwidth}
         \centering
         \includegraphics[width=\textwidth]{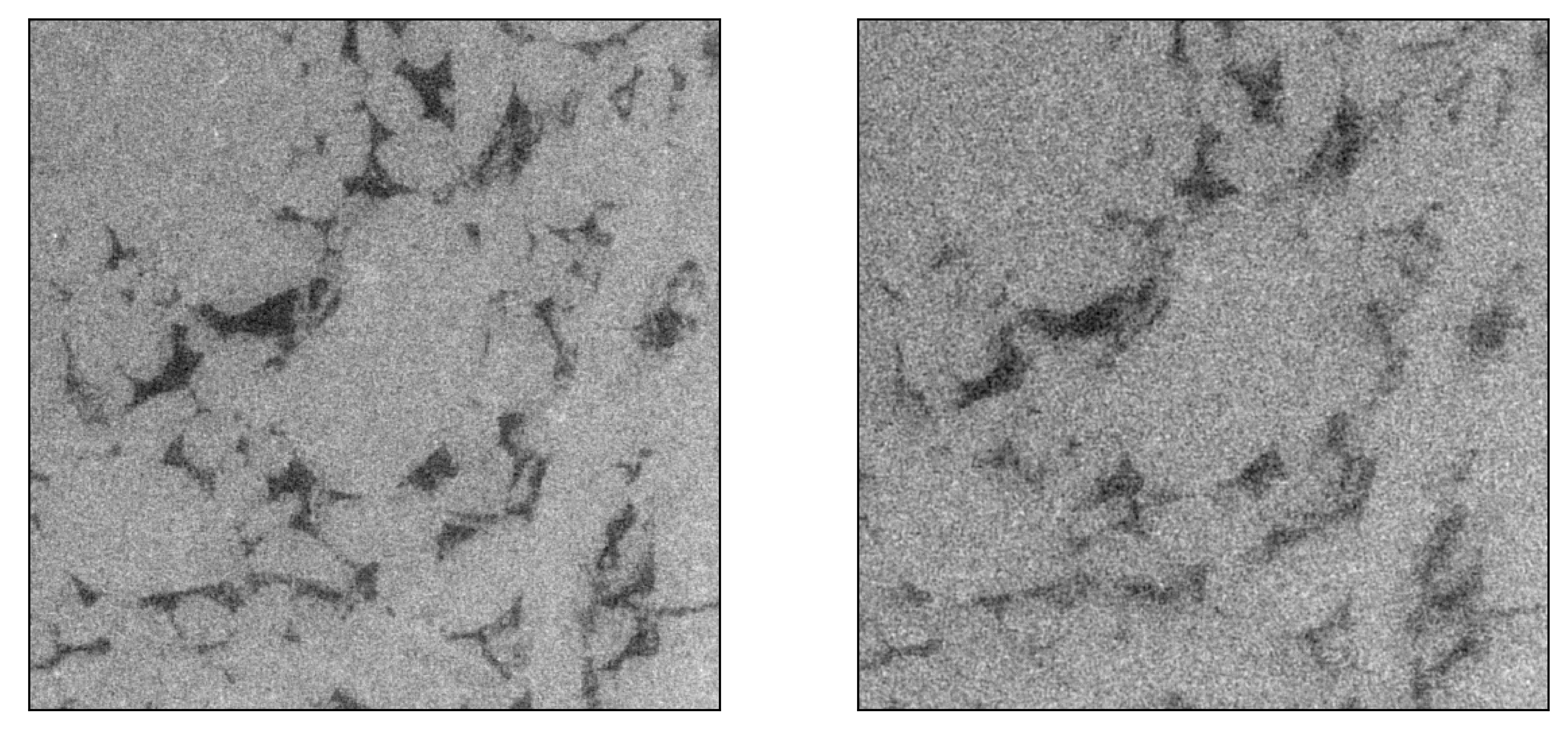}
         \label{fig:random_example}
         \caption{A pair of images from the test data set. The image on the left is an example of a high exposure time (1.4 s) scan, the image on the right is the equivalent low exposure time (0.5 s) scan (SSIM=0.52, PSNR=22 dB).}
         \label{fig: A random example from the test set}
     \end{subfigure}
     \begin{subfigure}[b]{\textwidth}
         \centering
         \includegraphics[width=\textwidth]{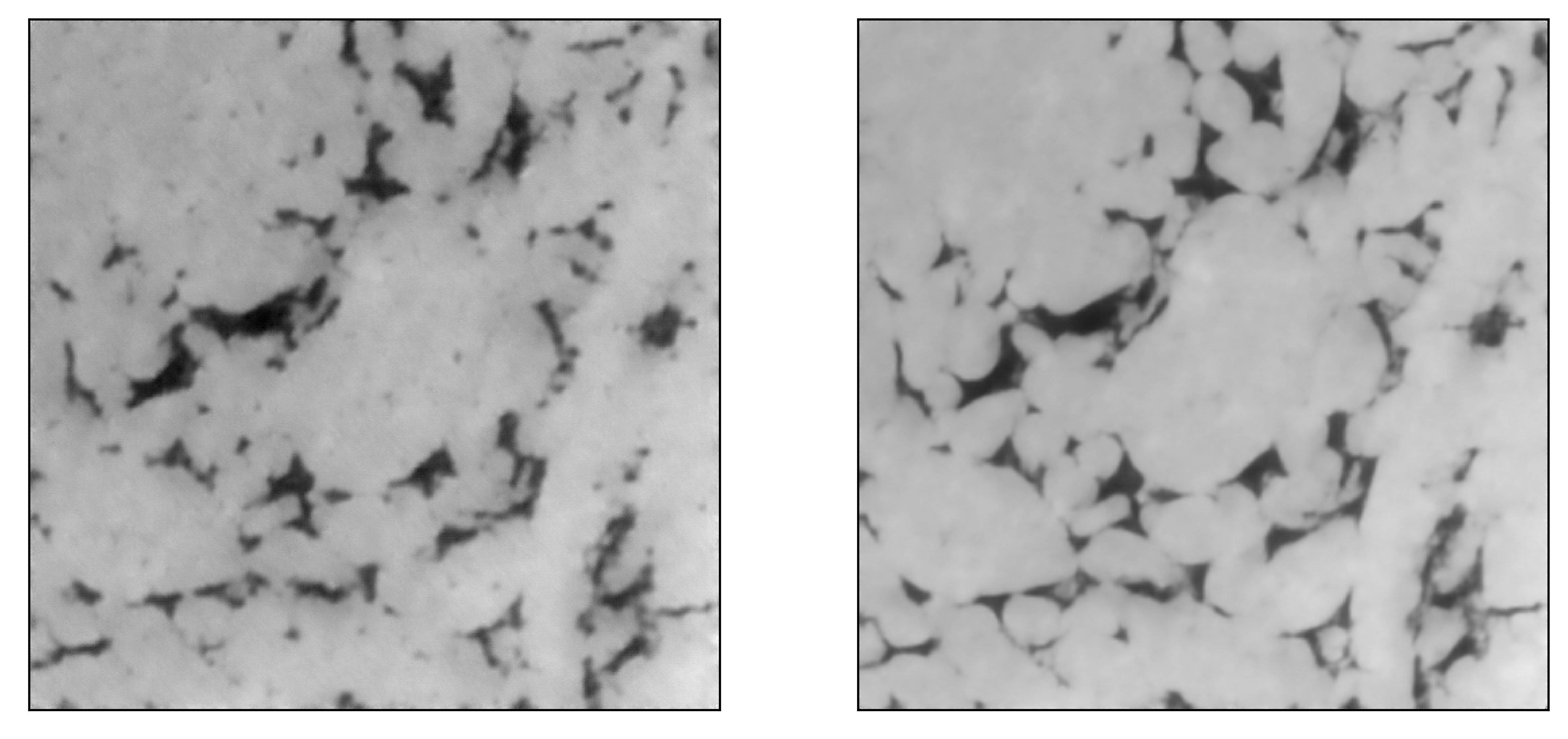}
         \caption{Denoising results exemplifying the performance of the DCNN (U-Net). The left image shows the prediction of the DCNN optimized with respect to the MSE (SSIM=0.77, PSNR=34 dB), the right image exhibits the prediction of the SSIM optimized network (SSIM=0.77, PSNR=34 dB). The MSE optimized network predicts coarser grain textures (greater variation in grayscale values indicative of larger variations in grain density) and boundaries and seems to be more sensitive to fine scale pore space (compare upper left quadrant of both images for the presence of fine scale pore space). Conversely, the SSIM optimized network suggests smoother textures, sharper grain boundaries and appears to be less sensitive to fine scale pore space.}
         \label{fig:denoising_example}
     \end{subfigure}
    \caption{DCNN (U-Net) denoising example from the test set.}
    \label{fig:denoising_random_example}
\end{figure}

Both loss functions perform remarkably well in restoring fine scale features, as exemplified in Figure \ref{fig:details_example}, and yield similar image quality improvements. With respect to Figure \ref{fig:denoising_random_example}, however, it seems they tend to emphasize different features of the data. The MSE optimized network predicts coarser grain textures and boundaries and seems to be more sensitive to fine scale pore space. Conversely, the SSIM optimized network suggests smoother textures, sharper grain boundaries and appears to be less sensitive to fine scale pore space.    

Surprisingly, the quality of the predicted image is clearly superior to the quality of the long exposure time image. As mentioned before, the network is seemingly able to distinguish between high frequency noise and discontinuities in form of edges. At this point it became necessary, to verify the predictive power of the networks and it was decided to create artificial cases where the ground truth is known. The approach is detailed in the next section.

\subsection{ASTRA Toolbox}
\label{astra_section}

\begin{figure}
    \includegraphics[width=\linewidth]{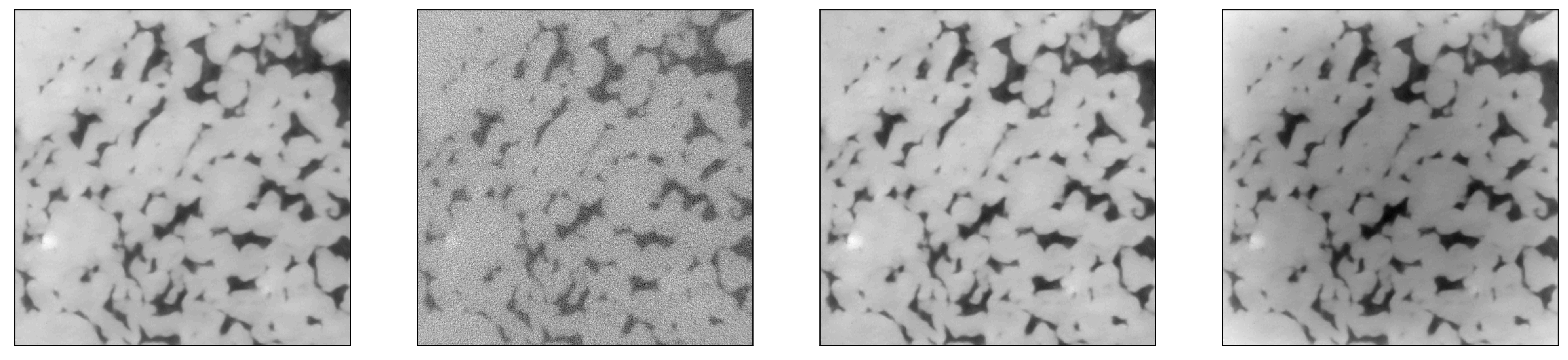}
  \caption{From left to right: Reference image (as predicted from the network representing the ground truth), FDK reconstruction, SIRT reconstruction, and CGLS reconstruction.}
  \label{fig:different_reconstructions}
\end{figure}

\begin{figure}
    \includegraphics[width=\linewidth]{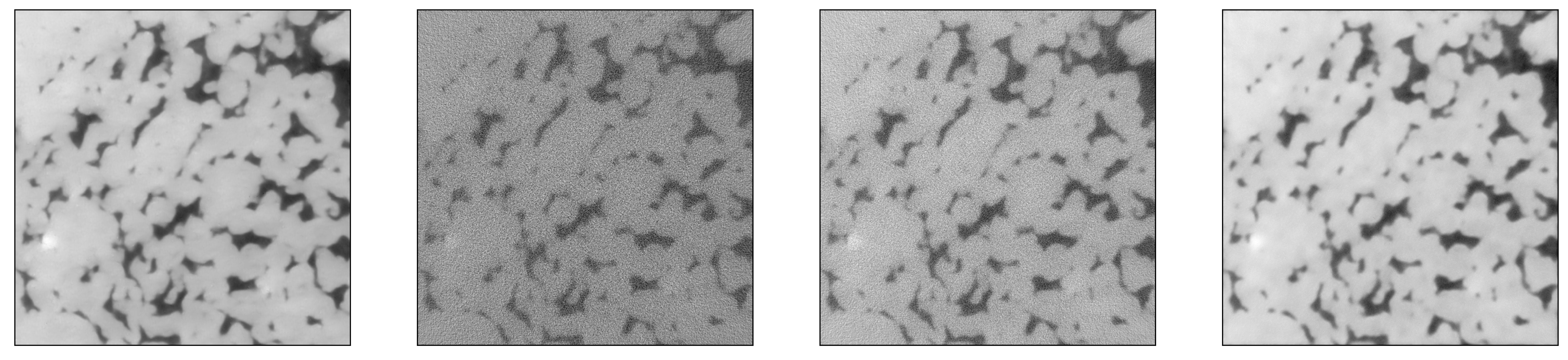}
  \caption{From left to right: Reference image (as predicted from the network representing the ground truth), artificial low exposure image created via FDK i.e. VDSR input (SSIM=0.17, PSNR=14 dB), artificial high exposure image created via FDK i.e. VDSR label (SSIM=0.30, PSNR=21 dB), and VDSR output (SSIM=0.89, PSNR=26 dB).}
  \label{fig:VDSR_denoising_for_artifical_data}
\end{figure}

As elaborated in the previous section, the images predicted by the DCNNs (images on the right of Figure \ref{fig:pre_trained_VDSR_example}, Figure \ref{fig:denoising_example}) are not only of superior quality compared to the low exposure images, but also exhibit less noise than their corresponding  high exposure images or the training labels.  As discussed in the introduction, and substantiated by Figure~\ref{fig:denoising_example}, noise can be reduced by increasing the exposure time or flux in general. In addition, the choice of the reconstruction algorithm is also critical. Iterative reconstruction algorithms like \emph{Simultaneous Iterative Reconstructive Technique} (SIRT) or \emph{Conjugate Gradient Least Squares} (CGLS) are well known to suppress noise compared to classic filtered backprojection (FDP) via Feldkamp-type (FDK) reconstruction algorithms \citep{fleischmann2011computed,biguri2016tigre}. The particular algorithm employed by the FEI Heliscan microCT is proprietary.

Given the surprising results, we seek to verify them by creating an artificial dataset for which the ground truth is known. For this purpose, the VDSR network's denoised images were fed into the ASTRA toolbox to create noisy projections mimicking low and high exposure time images. Next, the projections were reconstructed using FDK, SIRT, and CGLS. As summarized in Figure~\ref{fig:different_reconstructions}, SIRT and CGLS performed well in removing the noise whereas FDK failed to do so. Hence, we decided to solely focus on FDK for creation of the artificial data set. Subsequently, the artificial data set was tested utilizing the trained networks. 

Figure~\ref{fig:VDSR_denoising_for_artifical_data} shows results for the VDSR network trained on the artificial datasets i.e., it was trained to map the artificial low exposure to its corresponding artificial high exposure (training example/label). The average SSIM and PSNR values of the predicted images from the network (SSIM=0.86, PSNR=25 dB) are better than the artificial high (SSIM=0.28, PSNR=18 dB) and low exposure images (SSIM=0.17, PSNR=14 dB) according to 200 test images, where the reference images (ground truth) have been used to calculate these values. It is, again, surprising that the output images of the network yield greater quality results compared to their training examples (artificial high exposure images). This substantiates, however, the results reported in the previous section where the ground truth was unknown.

\section{Conclusions}
In this work, we have successfully demonstrated the value of DCNN to improve the quality of $\mu$-CT scans of a carbonte rock sample. The proposed method has the potential to reduce the exposure time by about 60\% (from 1.4 seconds to 0.5 seconds) without compromising the scan quality. On the contrary, we found that the networks are able to predict images of superior quality compared to the long exposure time training images (labels). In particular, the networks are seemingly able to distinguish between unwanted high frequency content like noise, and actual high frequency features of the data like discontinuities in form of grain boundaries. Importantly, we verified the predictive power of the networks by creating a synthetic dataset to compare against the known ground truth.

Given the substantial time requirements for training the networks, we also investigated the applicability of transfer learning. Using a pre-trained VDSR network we found that high quality images can be obtained for a smaller number of training epochs compared to training from scratch.

Additionally, we highlighted the impact of MSE and SSIM based loss functions on the DCNN predictions. Both yield similar improvements with respect to PSNR and SSIM. They tend to, however, emphasize different structural aspects of the specimen. The MSE optimized network predicts coarser grain textures and boundaries and seems to be more sensitive to fine scale pore space. Conversely, the SSIM optimized network suggests smoother textures, sharper grain boundaries and appears to be less sensitive to fine scale pore space.

To conclude, the proposed method enables substantial savings in acquisition time while simultaneously improving the scan quality. The reduction in scan time is an important aspect if dynamic processes are to be elucidated or higher sample throughput is required. Importantly, the approach is applicable to any computed tomography technology (medical CT, $\mu$-CT, industrial CT). The vast improvement in image quality, without the need for expert intervention, is crucial for digital rock physics applications where rock properties like porosity and permeability are estimated solely from computed tomography data. Inevitably, the accuracy of the estimation is dictated, in part, by the scan quality.

\medskip
\noindent
\section{Acknowledgements}
We thank Dr. Jack Dvorkin for approval to access the FEI Heliscan \mbox{microCT}, and Mr. Nadeem Ahmed Syed and Mr. Syed Rizwanullah Hussaini for explanations how to operate the scanner, all at the Center for Integrative Petroleum Research (CIPR)–CPG, KFUPM.

This work was supported by the Research Startup Grant no. SF20003 awarded to G.G. by the College of Petroleum Engineering and Geosciences, King Fahd University of Petroleum and Minerals.

\section{Conflict of Interest}

The authors declare no competing financial interest.

\bibliographystyle{elsarticle-harv}
\bibliography{CT_bib}

\begin{thebibliography}{44}
\expandafter\ifx\csname natexlab\endcsname\relax\def\natexlab#1{#1}\fi
\providecommand{\url}[1]{\texttt{#1}}
\providecommand{\href}[2]{#2}
\providecommand{\path}[1]{#1}
\providecommand{\DOIprefix}{doi:}
\providecommand{\ArXivprefix}{arXiv:}
\providecommand{\URLprefix}{URL: }
\providecommand{\Pubmedprefix}{pmid:}
\providecommand{\doi}[1]{\href{http://dx.doi.org/#1}{\path{#1}}}
\providecommand{\Pubmed}[1]{\href{pmid:#1}{\path{#1}}}
\providecommand{\bibinfo}[2]{#2}
\ifx\xfnm\relax \def\xfnm[#1]{\unskip,\space#1}\fi
\bibitem[{van Aarle et~al.(2016)van Aarle, Palenstijn, Cant, Janssens,
  Bleichrodt, Dabravolski, De~Beenhouwer, Batenburg and Sijbers}]{van2016fast}
\bibinfo{author}{van Aarle, W.}, \bibinfo{author}{Palenstijn, W.J.},
  \bibinfo{author}{Cant, J.}, \bibinfo{author}{Janssens, E.},
  \bibinfo{author}{Bleichrodt, F.}, \bibinfo{author}{Dabravolski, A.},
  \bibinfo{author}{De~Beenhouwer, J.}, \bibinfo{author}{Batenburg, K.J.},
  \bibinfo{author}{Sijbers, J.}, \bibinfo{year}{2016}.
\newblock \bibinfo{title}{Fast and flexible x-ray tomography using the astra
  toolbox}.
\newblock \bibinfo{journal}{Optics express} \bibinfo{volume}{24},
  \bibinfo{pages}{25129--25147}.
\bibitem[{van Aarle et~al.(2015)van Aarle, Palenstijn, {De Beenhouwer},
  Altantzis, Bals, Batenburg and Sijbers}]{VanAarle2015}
\bibinfo{author}{van Aarle, W.}, \bibinfo{author}{Palenstijn, W.J.},
  \bibinfo{author}{{De Beenhouwer}, J.}, \bibinfo{author}{Altantzis, T.},
  \bibinfo{author}{Bals, S.}, \bibinfo{author}{Batenburg, K.J.},
  \bibinfo{author}{Sijbers, J.}, \bibinfo{year}{2015}.
\newblock \bibinfo{title}{{The ASTRA Toolbox: A platform for advanced algorithm
  development in electron tomography}}.
\newblock \bibinfo{journal}{Ultramicroscopy} \bibinfo{volume}{157},
  \bibinfo{pages}{35--47}.
\newblock \DOIprefix\doi{10.1016/j.ultramic.2015.05.002}.
\bibitem[{Alqahtani et~al.(2020)Alqahtani, Alzubaidi, Armstrong, Swietojanski
  and Mostaghimi}]{alqahtani2020machine}
\bibinfo{author}{Alqahtani, N.}, \bibinfo{author}{Alzubaidi, F.},
  \bibinfo{author}{Armstrong, R.T.}, \bibinfo{author}{Swietojanski, P.},
  \bibinfo{author}{Mostaghimi, P.}, \bibinfo{year}{2020}.
\newblock \bibinfo{title}{Machine learning for predicting properties of porous
  media from 2d x-ray images}.
\newblock \bibinfo{journal}{Journal of Petroleum Science and Engineering}
  \bibinfo{volume}{184}, \bibinfo{pages}{106514}.
\bibitem[{Alshibli and Reed(2010)}]{Alshibli2010}
\bibinfo{author}{Alshibli, K.A.}, \bibinfo{author}{Reed, A.H.},
  \bibinfo{year}{2010}.
\newblock \bibinfo{title}{{Advances in Computed Tomography for Geomaterials}}.
\newblock \bibinfo{publisher}{John Wiley {\&} Sons, Inc.},
  \bibinfo{address}{Hoboken, NJ, USA}.
\newblock \DOIprefix\doi{10.1002/9781118557723}.
\bibitem[{Avşar and Arıca(2017)}]{Avsar2017}
\bibinfo{author}{Avşar, T.S.}, \bibinfo{author}{Arıca, S.},
  \bibinfo{year}{2017}.
\newblock \bibinfo{title}{{Automatic segmentation of computed tomography images
  of liver using watershed and thresholding algorithms}}, in:
  \bibinfo{booktitle}{IFMBE Proceedings}, \bibinfo{publisher}{Springer Verlag}.
  pp. \bibinfo{pages}{414--417}.
\newblock \DOIprefix\doi{10.1007/978-981-10-5122-7_104}.
\bibitem[{Barrett and Keat(2004)}]{barrett2004artifacts}
\bibinfo{author}{Barrett, J.F.}, \bibinfo{author}{Keat, N.},
  \bibinfo{year}{2004}.
\newblock \bibinfo{title}{Artifacts in ct: recognition and avoidance}.
\newblock \bibinfo{journal}{Radiographics} \bibinfo{volume}{24},
  \bibinfo{pages}{1679--1691}.
\bibitem[{Bartscher et~al.(2006)Bartscher, Hilpert, Goebbels, Weidemann, Puder
  and Jidav}]{Bartscher2006}
\bibinfo{author}{Bartscher, M.}, \bibinfo{author}{Hilpert, U.},
  \bibinfo{author}{Goebbels, J.}, \bibinfo{author}{Weidemann, G.},
  \bibinfo{author}{Puder, H.}, \bibinfo{author}{Jidav, H.N.},
  \bibinfo{year}{2006}.
\newblock \bibinfo{title}{{Einsatz von computer-tomographie in der
  Reverse-Engineering-Technologie}}.
\newblock \bibinfo{journal}{Materialpruefung/Materials Testing}
  \bibinfo{volume}{48}, \bibinfo{pages}{305--311}.
\newblock \DOIprefix\doi{10.3139/120.100208}.
\bibitem[{Bauer et~al.(2019)Bauer, Schrapp and Szijarto}]{Bauer2019}
\bibinfo{author}{Bauer, F.}, \bibinfo{author}{Schrapp, M.},
  \bibinfo{author}{Szijarto, J.}, \bibinfo{year}{2019}.
\newblock \bibinfo{title}{{Accuracy analysis of a piece-to-piece reverse
  engineering workflow for a turbine foil based on multi-modal computed
  tomography and additive manufacturing}}.
\newblock \bibinfo{journal}{Precision Engineering} \bibinfo{volume}{60},
  \bibinfo{pages}{63--75}.
\newblock \DOIprefix\doi{10.1016/j.precisioneng.2019.07.008}.
\bibitem[{Bazaikin et~al.(2017)Bazaikin, Gurevich, Iglauer, Khachkova,
  Kolyukhin, Lebedev, Lisitsa and Reshetova}]{bazaikin2017effect}
\bibinfo{author}{Bazaikin, Y.}, \bibinfo{author}{Gurevich, B.},
  \bibinfo{author}{Iglauer, S.}, \bibinfo{author}{Khachkova, T.},
  \bibinfo{author}{Kolyukhin, D.}, \bibinfo{author}{Lebedev, M.},
  \bibinfo{author}{Lisitsa, V.}, \bibinfo{author}{Reshetova, G.},
  \bibinfo{year}{2017}.
\newblock \bibinfo{title}{Effect of ct image size and resolution on the
  accuracy of rock property estimates}.
\newblock \bibinfo{journal}{Journal of Geophysical Research: Solid Earth}
  \bibinfo{volume}{122}, \bibinfo{pages}{3635--3647}.
\bibitem[{Berg et~al.(2017)Berg, Lopez and Berland}]{berg2017industrial}
\bibinfo{author}{Berg, C.F.}, \bibinfo{author}{Lopez, O.},
  \bibinfo{author}{Berland, H.}, \bibinfo{year}{2017}.
\newblock \bibinfo{title}{Industrial applications of digital rock technology}.
\newblock \bibinfo{journal}{Journal of Petroleum Science and Engineering}
  \bibinfo{volume}{157}, \bibinfo{pages}{131--147}.
\bibitem[{Biguri et~al.(2016)Biguri, Dosanjh, Hancock and
  Soleimani}]{biguri2016tigre}
\bibinfo{author}{Biguri, A.}, \bibinfo{author}{Dosanjh, M.},
  \bibinfo{author}{Hancock, S.}, \bibinfo{author}{Soleimani, M.},
  \bibinfo{year}{2016}.
\newblock \bibinfo{title}{Tigre: a matlab-gpu toolbox for cbct image
  reconstruction}.
\newblock \bibinfo{journal}{Biomedical Physics \& Engineering Express}
  \bibinfo{volume}{2}, \bibinfo{pages}{055010}.
\bibitem[{Boign{\'{e}} et~al.(2020)Boign{\'{e}}, Bennett, Wang, Mohri and
  Ihme}]{Boigne2020}
\bibinfo{author}{Boign{\'{e}}, E.}, \bibinfo{author}{Bennett, N.R.},
  \bibinfo{author}{Wang, A.}, \bibinfo{author}{Mohri, K.},
  \bibinfo{author}{Ihme, M.}, \bibinfo{year}{2020}.
\newblock \bibinfo{title}{{Simultaneous in-situ measurements of gas temperature
  and pyrolysis of biomass smoldering via X-ray computed tomography}}.
\newblock \bibinfo{journal}{Proceedings of the Combustion Institute}
  \DOIprefix\doi{10.1016/j.proci.2020.06.070}.
\bibitem[{Chen et~al.(2017a)Chen, Zhang, Kalra, Lin, Chen, Liao, Zhou and
  Wang}]{Chen2017}
\bibinfo{author}{Chen, H.}, \bibinfo{author}{Zhang, Y.},
  \bibinfo{author}{Kalra, M.K.}, \bibinfo{author}{Lin, F.},
  \bibinfo{author}{Chen, Y.}, \bibinfo{author}{Liao, P.},
  \bibinfo{author}{Zhou, J.}, \bibinfo{author}{Wang, G.},
  \bibinfo{year}{2017}a.
\newblock \bibinfo{title}{{Low-Dose CT with a residual encoder-decoder
  convolutional neural network}}.
\newblock \bibinfo{journal}{IEEE Transactions on Medical Imaging}
  \bibinfo{volume}{36}, \bibinfo{pages}{2524--2535}.
\newblock \DOIprefix\doi{10.1109/TMI.2017.2715284}.
\bibitem[{Chen et~al.(2017b)Chen, Zhang, Zhang, Liao, Li, Zhou and
  Wang}]{chen2017low}
\bibinfo{author}{Chen, H.}, \bibinfo{author}{Zhang, Y.},
  \bibinfo{author}{Zhang, W.}, \bibinfo{author}{Liao, P.}, \bibinfo{author}{Li,
  K.}, \bibinfo{author}{Zhou, J.}, \bibinfo{author}{Wang, G.},
  \bibinfo{year}{2017}b.
\newblock \bibinfo{title}{Low-dose ct denoising with convolutional neural
  network}, in: \bibinfo{booktitle}{2017 IEEE 14th International Symposium on
  Biomedical Imaging (ISBI 2017)}, \bibinfo{organization}{IEEE}. pp.
  \bibinfo{pages}{143--146}.
\bibitem[{Da~Wang et~al.(2019)Da~Wang, Armstrong and
  Mostaghimi}]{da2019enhancing}
\bibinfo{author}{Da~Wang, Y.}, \bibinfo{author}{Armstrong, R.T.},
  \bibinfo{author}{Mostaghimi, P.}, \bibinfo{year}{2019}.
\newblock \bibinfo{title}{Enhancing resolution of digital rock images with
  super resolution convolutional neural networks}.
\newblock \bibinfo{journal}{Journal of Petroleum Science and Engineering}
  \bibinfo{volume}{182}, \bibinfo{pages}{106261}.
\bibitem[{{De Chiffre} et~al.(2014){De Chiffre}, Carmignato, Kruth, Schmitt and
  Weckenmann}]{DeChiffre2014}
\bibinfo{author}{{De Chiffre}, L.}, \bibinfo{author}{Carmignato, S.},
  \bibinfo{author}{Kruth, J.P.}, \bibinfo{author}{Schmitt, R.},
  \bibinfo{author}{Weckenmann, A.}, \bibinfo{year}{2014}.
\newblock \bibinfo{title}{{Industrial applications of computed tomography}}.
\newblock \bibinfo{journal}{CIRP Annals - Manufacturing Technology}
  \bibinfo{volume}{63}, \bibinfo{pages}{655--677}.
\newblock \DOIprefix\doi{10.1016/j.cirp.2014.05.011}.
\bibitem[{Diwakar and Kumar(2018)}]{Diwakar2018}
\bibinfo{author}{Diwakar, M.}, \bibinfo{author}{Kumar, M.},
  \bibinfo{year}{2018}.
\newblock \bibinfo{title}{{A review on CT image noise and its denoising}}.
\newblock \DOIprefix\doi{10.1016/j.bspc.2018.01.010}.
\bibitem[{Feldkamp et~al.(1984)Feldkamp, Davis and Kress}]{Feldkamp1984}
\bibinfo{author}{Feldkamp, L.A.}, \bibinfo{author}{Davis, L.C.},
  \bibinfo{author}{Kress, J.W.}, \bibinfo{year}{1984}.
\newblock \bibinfo{title}{{Practical cone-beam algorithm}}.
\newblock \bibinfo{journal}{Journal of the Optical Society of America A}
  \bibinfo{volume}{1}, \bibinfo{pages}{612}.
\newblock \DOIprefix\doi{10.1364/josaa.1.000612}.
\bibitem[{Fleischmann and Boas(2011)}]{fleischmann2011computed}
\bibinfo{author}{Fleischmann, D.}, \bibinfo{author}{Boas, F.E.},
  \bibinfo{year}{2011}.
\newblock \bibinfo{title}{Computed tomography—old ideas and new technology}.
\bibitem[{Frommer and Maass(1999)}]{Frommer1999}
\bibinfo{author}{Frommer, A.}, \bibinfo{author}{Maass, P.},
  \bibinfo{year}{1999}.
\newblock \bibinfo{title}{{Fast CG-based methods for Tikhonov-Phillips
  regularization}}.
\newblock \bibinfo{journal}{SIAM Journal of Scientific Computing}
  \bibinfo{volume}{20}, \bibinfo{pages}{1831--1850}.
\newblock \DOIprefix\doi{10.1137/S1064827596313310}.
\bibitem[{Gilbert(1972)}]{Gilbert1972}
\bibinfo{author}{Gilbert, P.}, \bibinfo{year}{1972}.
\newblock \bibinfo{title}{{Iterative methods for the three-dimensional
  reconstruction of an object from projections}}.
\newblock \bibinfo{journal}{Journal of Theoretical Biology}
  \bibinfo{volume}{36}, \bibinfo{pages}{105--117}.
\newblock \DOIprefix\doi{10.1016/0022-5193(72)90180-4}.
\bibitem[{Glatz et~al.(2016)Glatz, Castanier and Kovscek}]{Glatz2016}
\bibinfo{author}{Glatz, G.}, \bibinfo{author}{Castanier, L.},
  \bibinfo{author}{Kovscek, A.}, \bibinfo{year}{2016}.
\newblock \bibinfo{title}{{Visualization and Quantification of Thermally
  Induced Porosity Alteration of Immature Source Rock Using X-ray Computed
  Tomography}}.
\newblock \bibinfo{journal}{Energy and Fuels} \bibinfo{volume}{30}.
\newblock \DOIprefix\doi{10.1021/acs.energyfuels.6b01430}.
\bibitem[{Glatz et~al.(2018)Glatz, Lapene, Castanier and Kovscek}]{Glatz2018}
\bibinfo{author}{Glatz, G.}, \bibinfo{author}{Lapene, A.},
  \bibinfo{author}{Castanier, L.M.}, \bibinfo{author}{Kovscek, A.R.},
  \bibinfo{year}{2018}.
\newblock \bibinfo{title}{{An experimental platform for triaxial
  high-pressure/high-temperature testing of rocks using computed tomography}}.
\newblock \bibinfo{journal}{Review of Scientific Instruments}
  \bibinfo{volume}{89}, \bibinfo{pages}{45101}.
\newblock \DOIprefix\doi{10.1063/1.5030204}.
\bibitem[{Goldman(2007)}]{goldman2007principles}
\bibinfo{author}{Goldman, L.W.}, \bibinfo{year}{2007}.
\newblock \bibinfo{title}{Principles of ct: radiation dose and image quality}.
\newblock \bibinfo{journal}{Journal of nuclear medicine technology}
  \bibinfo{volume}{35}, \bibinfo{pages}{213--225}.
\bibitem[{Gravel et~al.(2004)Gravel, Beaudoin and {De Guise}}]{Gravel2004}
\bibinfo{author}{Gravel, P.}, \bibinfo{author}{Beaudoin, G.},
  \bibinfo{author}{{De Guise}, J.A.}, \bibinfo{year}{2004}.
\newblock \bibinfo{title}{{A method for modeling noise in medical images}}.
\newblock \bibinfo{journal}{IEEE Transactions on Medical Imaging}
  \bibinfo{volume}{23}, \bibinfo{pages}{1221--1232}.
\newblock \DOIprefix\doi{10.1109/TMI.2004.832656}.
\bibitem[{Guan et~al.(2019)Guan, Nazarova, Guo, Tchelepi, Kovscek and
  Creux}]{guan2019effects}
\bibinfo{author}{Guan, K.M.}, \bibinfo{author}{Nazarova, M.},
  \bibinfo{author}{Guo, B.}, \bibinfo{author}{Tchelepi, H.},
  \bibinfo{author}{Kovscek, A.R.}, \bibinfo{author}{Creux, P.},
  \bibinfo{year}{2019}.
\newblock \bibinfo{title}{Effects of image resolution on sandstone porosity and
  permeability as obtained from x-ray microscopy}.
\newblock \bibinfo{journal}{Transport in Porous Media} \bibinfo{volume}{127},
  \bibinfo{pages}{233--245}.
\bibitem[{He et~al.(2015)He, Zhang, Ren and Sun}]{DBLP:journals/corr/HeZR015}
\bibinfo{author}{He, K.}, \bibinfo{author}{Zhang, X.}, \bibinfo{author}{Ren,
  S.}, \bibinfo{author}{Sun, J.}, \bibinfo{year}{2015}.
\newblock \bibinfo{title}{Delving deep into rectifiers: Surpassing human-level
  performance on imagenet classification}.
\newblock \bibinfo{journal}{CoRR} \bibinfo{volume}{abs/1502.01852}.
\newblock \URLprefix \url{http://arxiv.org/abs/1502.01852},
  \href{http://arxiv.org/abs/1502.01852}{{\tt arXiv:1502.01852}}.
\bibitem[{He et~al.(2014)He, Zhang and Lu}]{He2014}
\bibinfo{author}{He, N.}, \bibinfo{author}{Zhang, L.}, \bibinfo{author}{Lu,
  K.}, \bibinfo{year}{2014}.
\newblock \bibinfo{title}{{Aluminum CT image defect detection based on
  segmentation and feature extraction}}, in: \bibinfo{booktitle}{Lecture Notes
  in Computer Science (including subseries Lecture Notes in Artificial
  Intelligence and Lecture Notes in Bioinformatics)},
  \bibinfo{publisher}{Springer Verlag}. pp. \bibinfo{pages}{446--454}.
\newblock \DOIprefix\doi{10.1007/978-3-319-07626-3_41}.
\bibitem[{Kang et~al.(2017)Kang, Min and Ye}]{Kang2017}
\bibinfo{author}{Kang, E.}, \bibinfo{author}{Min, J.}, \bibinfo{author}{Ye,
  J.C.}, \bibinfo{year}{2017}.
\newblock \bibinfo{title}{{A deep convolutional neural network using
  directional wavelets for low-dose X-ray CT reconstruction}}.
\newblock \bibinfo{journal}{Medical Physics} \bibinfo{volume}{44},
  \bibinfo{pages}{e360--e375}.
\newblock \DOIprefix\doi{10.1002/mp.12344},
  \href{http://arxiv.org/abs/1610.09736}{{\tt arXiv:1610.09736}}.
\bibitem[{Kim et~al.(2016)Kim, Kwon~Lee and Mu~Lee}]{kim2016accurate}
\bibinfo{author}{Kim, J.}, \bibinfo{author}{Kwon~Lee, J.},
  \bibinfo{author}{Mu~Lee, K.}, \bibinfo{year}{2016}.
\newblock \bibinfo{title}{Accurate image super-resolution using very deep
  convolutional networks}, in: \bibinfo{booktitle}{Proceedings of the IEEE
  conference on computer vision and pattern recognition}, pp.
  \bibinfo{pages}{1646--1654}.
\bibitem[{Lee(1981)}]{Lee1981}
\bibinfo{author}{Lee, J.S.}, \bibinfo{year}{1981}.
\newblock \bibinfo{title}{{Refined filtering of image noise using local
  statistics}}.
\newblock \bibinfo{journal}{Computer Graphics and Image Processing}
  \bibinfo{volume}{15}, \bibinfo{pages}{380--389}.
\newblock \DOIprefix\doi{10.1016/S0146-664X(81)80018-4}.
\bibitem[{Liu et~al.(2018)Liu, Jin and Wang}]{liu2018critical}
\bibinfo{author}{Liu, T.}, \bibinfo{author}{Jin, X.}, \bibinfo{author}{Wang,
  M.}, \bibinfo{year}{2018}.
\newblock \bibinfo{title}{Critical resolution and sample size of digital rock
  analysis for unconventional reservoirs}.
\newblock \bibinfo{journal}{Energies} \bibinfo{volume}{11},
  \bibinfo{pages}{1798}.
\bibitem[{Macovski(1983)}]{macovski_medical_1983}
\bibinfo{author}{Macovski, A.}, \bibinfo{year}{1983}.
\newblock \bibinfo{title}{Medical Imaging Systems}.
\newblock \bibinfo{publisher}{Prentice-Hall}.
\bibitem[{MathWorks(2018)}]{mathworks}
\bibinfo{author}{MathWorks}, \bibinfo{year}{2018}.
\newblock \bibinfo{title}{Single image super resolution using deep learning}.
\newblock
  \bibinfo{note}{\url{https://www.mathworks.com/help/images/single-image-super-resolution-using-deep-learning.html},
  Accessed: 4 May 2020}.
\bibitem[{McCollough et~al.(2009)McCollough, Primak, Braun, Kofler, Yu and
  Christner}]{mccollough2009strategies}
\bibinfo{author}{McCollough, C.H.}, \bibinfo{author}{Primak, A.N.},
  \bibinfo{author}{Braun, N.}, \bibinfo{author}{Kofler, J.},
  \bibinfo{author}{Yu, L.}, \bibinfo{author}{Christner, J.},
  \bibinfo{year}{2009}.
\newblock \bibinfo{title}{Strategies for reducing radiation dose in ct}.
\newblock \bibinfo{journal}{Radiologic Clinics} \bibinfo{volume}{47},
  \bibinfo{pages}{27--40}.
\bibitem[{Mostaghimi et~al.(2013)Mostaghimi, Blunt and
  Bijeljic}]{Mostaghimi2013}
\bibinfo{author}{Mostaghimi, P.}, \bibinfo{author}{Blunt, M.J.},
  \bibinfo{author}{Bijeljic, B.}, \bibinfo{year}{2013}.
\newblock \bibinfo{title}{{Computations of Absolute Permeability on Micro-CT
  Images}}.
\newblock \bibinfo{journal}{Mathematical Geosciences} \bibinfo{volume}{45},
  \bibinfo{pages}{103--125}.
\newblock \DOIprefix\doi{10.1007/s11004-012-9431-4}.
\bibitem[{Nishio et~al.(2017)Nishio, Nagashima, Hirabayashi, Ohnishi, Sasaki,
  Sagawa, Hamada and Yamashita}]{nishio2017convolutional}
\bibinfo{author}{Nishio, M.}, \bibinfo{author}{Nagashima, C.},
  \bibinfo{author}{Hirabayashi, S.}, \bibinfo{author}{Ohnishi, A.},
  \bibinfo{author}{Sasaki, K.}, \bibinfo{author}{Sagawa, T.},
  \bibinfo{author}{Hamada, M.}, \bibinfo{author}{Yamashita, T.},
  \bibinfo{year}{2017}.
\newblock \bibinfo{title}{Convolutional auto-encoder for image denoising of
  ultra-low-dose ct}.
\newblock \bibinfo{journal}{Heliyon} \bibinfo{volume}{3},
  \bibinfo{pages}{e00393}.
\bibitem[{Papari et~al.(2016)Papari, Idowu and Varslot}]{papari2016fast}
\bibinfo{author}{Papari, G.}, \bibinfo{author}{Idowu, N.},
  \bibinfo{author}{Varslot, T.}, \bibinfo{year}{2016}.
\newblock \bibinfo{title}{Fast bilateral filtering for denoising large 3d
  images}.
\newblock \bibinfo{journal}{Ieee transactions on image processing}
  \bibinfo{volume}{26}, \bibinfo{pages}{251--261}.
\bibitem[{Pelt et~al.(2018)Pelt, Batenburg and Sethian}]{pelt2018improving}
\bibinfo{author}{Pelt, D.M.}, \bibinfo{author}{Batenburg, K.J.},
  \bibinfo{author}{Sethian, J.A.}, \bibinfo{year}{2018}.
\newblock \bibinfo{title}{Improving tomographic reconstruction from limited
  data using mixed-scale dense convolutional neural networks}.
\newblock \bibinfo{journal}{Journal of Imaging} \bibinfo{volume}{4},
  \bibinfo{pages}{128}.
\bibitem[{du~Plessis et~al.(2016)du~Plessis, le~Roux and
  Guelpa}]{DuPlessis2016}
\bibinfo{author}{du~Plessis, A.}, \bibinfo{author}{le~Roux, S.G.},
  \bibinfo{author}{Guelpa, A.}, \bibinfo{year}{2016}.
\newblock \bibinfo{title}{{Comparison of medical and industrial X-ray computed
  tomography for non-destructive testing}}.
\newblock \bibinfo{journal}{Case Studies in Nondestructive Testing and
  Evaluation} \bibinfo{volume}{6}, \bibinfo{pages}{17--25}.
\newblock \DOIprefix\doi{10.1016/j.csndt.2016.07.001}.
\bibitem[{Wang et~al.(2019)Wang, Teng, He, Feng and Zhang}]{wang2019ct}
\bibinfo{author}{Wang, Y.}, \bibinfo{author}{Teng, Q.}, \bibinfo{author}{He,
  X.}, \bibinfo{author}{Feng, J.}, \bibinfo{author}{Zhang, T.},
  \bibinfo{year}{2019}.
\newblock \bibinfo{title}{Ct-image of rock samples super resolution using 3d
  convolutional neural network}.
\newblock \bibinfo{journal}{Computers \& Geosciences} \bibinfo{volume}{133},
  \bibinfo{pages}{104314}.
\bibitem[{Wang et~al.(2004)Wang, Bovik, Sheikh and Simoncelli}]{wang2004image}
\bibinfo{author}{Wang, Z.}, \bibinfo{author}{Bovik, A.C.},
  \bibinfo{author}{Sheikh, H.R.}, \bibinfo{author}{Simoncelli, E.P.},
  \bibinfo{year}{2004}.
\newblock \bibinfo{title}{Image quality assessment: from error visibility to
  structural similarity}.
\newblock \bibinfo{journal}{IEEE transactions on image processing}
  \bibinfo{volume}{13}, \bibinfo{pages}{600--612}.
\bibitem[{Willemink and No{\"e}l(2019)}]{willemink2019evolution}
\bibinfo{author}{Willemink, M.J.}, \bibinfo{author}{No{\"e}l, P.B.},
  \bibinfo{year}{2019}.
\newblock \bibinfo{title}{The evolution of image reconstruction for ct—from
  filtered back projection to artificial intelligence}.
\newblock \bibinfo{journal}{European radiology} \bibinfo{volume}{29},
  \bibinfo{pages}{2185--2195}.
\bibitem[{Zhao et~al.(2016)Zhao, Gallo, Frosio and Kautz}]{zhao2016loss}
\bibinfo{author}{Zhao, H.}, \bibinfo{author}{Gallo, O.},
  \bibinfo{author}{Frosio, I.}, \bibinfo{author}{Kautz, J.},
  \bibinfo{year}{2016}.
\newblock \bibinfo{title}{Loss functions for image restoration with neural
  networks}.
\newblock \bibinfo{journal}{IEEE Transactions on computational imaging}
  \bibinfo{volume}{3}, \bibinfo{pages}{47--57}.

\end{thebibliography}
\end{document}